\documentclass[12pt, a4paper]{article}

\usepackage{rotating}
\usepackage{amsmath}
\usepackage{mathtools}
\usepackage{amsthm}
\usepackage{amsfonts}
\usepackage[dvips]{epsfig}
\usepackage{graphicx}
\usepackage{amssymb}
\usepackage{cancel}
\usepackage{pstricks} 
\usepackage{lscape}
\usepackage{cancel}
\usepackage{slashed}
\usepackage{pstricks} 
\usepackage{lscape}
\usepackage{color}
\usepackage{cite}
\newcommand{\beq}{\begin{equation}}
\newcommand{\eeq}{\end{equation}}
\newcommand{\ga}{\lower.7ex\hbox{$\;\stackrel{\textstyle>}{\sim}\;$}}
\newcommand{\la}{\lower.7ex\hbox{$\;\stackrel{\textstyle<}{\sim}\;$}}

\newcommand{\kahler}{K\"ahler }

\begin{document}

\begin{flushright}
{\tt KCL-PH-TH/2016-54}, {\tt CERN-PH-TH/2016-188}  \\
{\tt UT-16-29, ACT-05-16, MI-TH-1625} \\
{\tt UMN-TH-3603/16, FTPI-MINN-16/26} \\
\end{flushright}

\vspace{0.7cm}
\begin{center}
{\bf {\large Starobinsky-Like Inflation and Neutrino Masses \\
\vspace{0.1in}
in a No-Scale SO(10) Model}}
\vspace {0.1in}
\end{center}

\vspace{0.05in}

\begin{center}{
{\bf John~Ellis}$^{a}$,
{\bf Marcos~A.~G.~Garcia}$^{b}$,
{\bf Natsumi Nagata}$^{c}$, \\
{\bf Dimitri~V.~Nanopoulos}$^{d}$ and
{\bf Keith~A.~Olive}$^{e}$
}
\end{center}

\begin{center}
{\em $^a$Theoretical Particle Physics and Cosmology Group, Department of
  Physics, King's~College~London, London WC2R 2LS, United Kingdom;\\
Theory Division, CERN, CH-1211 Geneva 23,
  Switzerland}\\[0.2cm]
  {\em $^b$Physics \& Astronomy Department, Rice University, Houston, TX 77005, USA}\\[0.2cm]
  {\em $^c$Department of Physics, University of Tokyo, Bunkyo-ku, Tokyo 113--0033, Japan}\\[0.2cm]
{\em $^d$George P. and Cynthia W. Mitchell Institute for Fundamental Physics and Astronomy,
Texas A\&M University, College Station, TX 77843, USA;\\
Astroparticle Physics Group, Houston Advanced Research Center (HARC), \\ Mitchell Campus, Woodlands, TX 77381, USA;\\
Academy of Athens, Division of Natural Sciences,
Athens 10679, Greece}\\[0.2cm]
  {\em $^e$William I. Fine Theoretical Physics Institute, School of Physics and Astronomy,\\
University of Minnesota, Minneapolis, MN 55455, USA}
 
\end{center}

\bigskip

\centerline{\bf ABSTRACT}

\noindent  
{\small Using a no-scale supergravity framework, we construct an SO(10)
model that makes predictions for cosmic microwave background observables
similar to those of the Starobinsky model of inflation, and incorporates
a double-seesaw model for neutrino masses consistent with oscillation
experiments and late-time cosmology. We pay particular attention to the
behaviour of the scalar fields during inflation and the subsequent
reheating. }

\vspace{0.2in}

\begin{flushleft}
September 2016
\end{flushleft}
\medskip
\noindent

\newpage

\section{Introduction}

Recent measurements of the cosmic microwave background (CMB)
\cite{planck15,planckbicep} provide important constraints on the scalar
tilt $n_s$ and tensor-to-scalar ratio $r$ in the perturbation spectrum,
which in turn provide important restrictions on possible models of
cosmological inflation \cite{reviews}. Among the models that fit the
data very well is the Starobinsky model \cite{Staro,MC,Staro2} that is
based on an $R + R^2$ modification of minimal Einstein gravity. Another
model that is consistent with the CMB data is Higgs inflation
\cite{Higgsinf}, which assumes a non-minimal coupling of the Standard
Model Higgs field to gravity\footnote{This model is disfavoured by
current measurements of the top and Higgs masses, which indicate that
the effective Higgs potential becomes negative at large field
values~\cite{Buttazzo}, unless the Standard Model is supplemented by new
physics.}. A central challenge in inflationary model-building is
therefore the construction of a model that incorporates not only the
Standard Model but also plausible candidates for new physics beyond,
such as neutrino masses and oscillations, dark matter, and the baryon
asymmetry of the Universe. 

Among the leading frameworks for physics beyond the Standard Model at
the TeV scale and above is supersymmetry. It has many advantages for
particle physics, could provide the astrophysical dark matter, offers
new mechanisms for generating the baryon asymmetry, and could also
stabilize the small potential parameters required in generic models of
inflation \cite{ENOT}. In cosmological applications, it is essential to
combine supersymmetry with gravity in the supergravity
framework~\cite{nost,hrr,gl1}. However, generic supergravity models are
not suitable for cosmology, since their effective potentials contain
`holes' of depth ${\cal O}(1)$ in natural units \cite{eta}, an obstacle
known as the $\eta$ problem. One exception to this `holy' rule is
provided by no-scale supergravity \cite{no-scale,LN}, which offers an
effective potential that is positive semi-definite at the tree level,
and has the added motivation that it appears in compactifications of
string theory \cite{Witten1985}. In this case, the $\eta$ problem can be
avoided \cite{GMO} and it is natural, therefore, to consider
inflationary models in this context \cite{gl2,KQ,EENOS,otherns}.

Consequently \cite{ENO6}, there has been continuing interest in
constructing no-scale supergravity models of inflation
\cite{ENO7,ENO8,KL,others,EGNO4,EGNO5,EGNO6,GM}, which lead naturally to
predictions for the CMB variables $(n_s, r)$ that are similar to those
of the Starobinsky model~\cite{ENO6}. In particular, no-scale models
have been constructed in which the inflaton could be identified with a
singlet (right-handed) sneutrino \cite{ENO8,EGNO4,EGNO5}, and also
no-scale GUT models have been constructed in which the inflaton is
identified with a supersymmetric Higgs boson, avoiding the problems of
conventional Higgs inflation \cite{sHiggs}.

In this paper we take an alternative approach to the construction of a
no-scale GUT model of inflation, namely we consider a supersymmetric
SO(10) GUT in which the sneutrino is embedded in a $\mathbf{16}$ of the
gauge group and the inflaton is identified with a singlet of SO(10). We
show that this model also makes Starobinsky-like predictions for the CMB
variables $(n_s, r)$. However, achieving this result makes non-trivial
demands on the structure of the SO(10) model, which we study in this
paper. 

One issue is the behaviour of the GUT non-singlet scalar fields during
inflation, which we require to be such that the model predictions are
Starobinsky-like. Another issue is the form of the neutrino mass
matrix. In our model, the superpartner of the inflaton field mixes with
the doublet (left-handed) and singlet (right-handed) neutrino fields,
leading to a double-seesaw structure, which must satisfy certain
conditions if it is to give masses for the light (mainly left-handed)
neutrinos that are compatible with oscillation experiments and late-time
cosmology. Finally, we also consider the issue of reheating
and the generation of the baryon asymmetry following
inflation, which, in addition to being compatible with the Planck
constraints on $n_s$, should not lead to overproduction of gravitinos. 

We find parameters for the no-scale SO(10) GUT model that are compatible
with all these cosmological and neutrino constraints, providing an
existence proof for a more complete model of particle physics and
cosmology than has been provided in previous Starobinsky-like no-scale
supergravity models of inflation. 

The structure of this paper is as follows. In Section~\ref{sec:setup} we
set up our inflationary model, including the no-scale and SO(10) aspects
of its framework. The realization of inflation in this model is
described in Section~\ref{inflation}, paying particular attention to the
requirements that its predictions resemble those of the Starobinsky
model. Section~\ref{sec:neutrinomass} explores the generation of
neutrino masses in this model, as they are generated via a double-seesaw
mechanism. Reheating and leptogenesis after inflation is discussed in
Section~\ref{sec:reheating}, with particular attention paid to the
gravitino abundance. Finally, our conclusions are summarized in
Section~\ref{sec:summary}.

\section{An SO(10) Inflationary Model Set-Up in No-Scale Supergravity}
\label{sec:setup}

\subsection{No-scale framework}

No-scale supergravity provides a remarkably simple field-theoretic realization
of predictions for the CMB observables that are similar to those of the
$R + R^2$ Starobinsky model of inflation \cite{ENO6}. In the minimal
two-field case~\cite{EKNN} useful for inflation, the K\"ahler potential
can be written as
\begin{equation}
K \; \ni \; - 3 \, \ln \left(T + T^* - \frac{|\phi|^2}{3} + \dots \right) + \dots \, , 
\label{K21}
\end{equation}
where $T$ and $\phi$ are complex scalar fields and the $\dots$ represent
possible additional matter fields; untwisted if in the log, twisted if outside.
Restricting our attention to the the two-field case for the moment,
the K\"ahler potential (\ref{K21}) yields the following kinetic terms
for the scalar fields $T$ and $\phi$:
\begin{equation}
{\cal L}_{\rm KE} =   
\frac{3}{(T + T^* - |\phi|^2/3)^2} ~
\left( \partial_\mu \phi^*, \partial_\mu T^* \right)
\begin{pmatrix} 
(T + T^*)/3 & - \phi \\ - \phi^* & 1 
\end{pmatrix}
\begin{pmatrix} 
\partial^\mu \phi \\ \partial^\mu T 
\end{pmatrix}  \, .
\label{no-scaleL}
\end{equation}
For a general superpotential $W(T,\phi)$, the
effective potential becomes
\begin{equation}
V \; = \; \frac{{\hat V}}{(T + T^* - |\phi|^2/3)^2}  ~,
\label{VVhatT}
\eeq
with
\beq
{\hat V} \; \equiv \; \left| \frac{\partial W}{\partial \phi} \right|^2  +\frac{1}{3} (T+T^*) |W_T|^2 +
\frac{1}{3} \left(W_T (\phi^* W_\phi^* - 3 W^*) + {\rm h.c.}  \right) \, ,
\label{effV}
\end{equation}
where $W_\phi \equiv \partial W/\partial \phi$ and $W_T \equiv \partial
W/\partial T$. 

If the modulus $T$ is fixed with a vacuum expectation value (vev) $2
\,\langle {\rm Re} T \rangle = c$ and $\langle {\rm Im} T \rangle = 0$,
as was shown in~Ref.~\cite{ENO6}, the Starobinsky inflationary potential
\begin{equation} 
V =  \frac{3}{4} M^2 (1- e^{-\sqrt{2/3}\phi^\prime})^2
\label{r2pot}
\end{equation}
would be obtained
with the following Wess--Zumino choice of superpotential~\cite{CEM}:
\begin{equation}
W \; = \; \frac{\hat \mu}{2} \phi^2 - \frac{\lambda}{3} \phi^3 \, 
\label{WZW}
\end{equation}
if $\lambda = \mu/3$ where $\mu = {\hat \mu} / \sqrt{c/3}$, as may be
seen after a field redefinition to a canonically-normalized inflaton
field $\phi'$. In order to obtain the correct amplitude for density
fluctuations, we must take $M = \mu/\sqrt{3} \approx 10^{-5}$ in natural
units with $M_P^{-2} = 8 \pi G_N \equiv 1$. Alternatively, if the field
$\phi$ is fixed (with $\phi = 0$), and the superpotential is given by
\cite{Cecotti} 
\begin{equation}
W \; = \sqrt{3} M \phi (T - 1/2) \, ,
\label{W3}
\end{equation}
the Starobinsky potential (\ref{r2pot}) is found when $T$ is converted
to a canonical field. In fact, there is a large class of superpotentials
that all lead to the same inflationary potential \cite{ENO7}. The
stabilization of either $\phi$ or $T$ in this context can be achieved
through quartic terms in the K\"ahler potential
\cite{EKN3,ENO7,KL,EGNO1,EGNO4,EGNO5,EGNO6}.

In order to achieve reheating the inflaton must be coupled  to matter. 
In no-scale models, supergravity couplings of the inflaton are strongly
suppressed \cite{ekoty}, and require either a non-trivial coupling
through the gauge kinetic function \cite{ekoty,klor,EGNO5}, or a direct
coupling to the matter sector through the superpotential. It was
proposed in Ref.~\cite{ENO8} that the inflaton could be associated with
the scalar component of the right-handed (SU(2)-singlet) neutrino
superfield $\nu_R$, and a specific no-scale supersymmetric GUT
\cite{superGUT,Ellis:2016tjc} model based on SU(5) was proposed, in
which the $\nu_R$ appeared as a singlet. In this model, reheating takes
place when the inflaton decays into the left-handed sneutrino and Higgs
(or neutrino and Higgsino), and may occur simultaneously with
leptogenesis \cite{lepto}.

\subsection{SO(10) GUT Construction}

We consider here possibilities for no-scale inflation in the context of
SO(10) grand unification \cite{so10,GN}. We immediately observe that, if we
consider the superpotential \eqref{WZW} for the inflaton, then $\phi$
cannot be associated with the right-handed neutrino. This is because, in
SO(10), the $\nu_R$ is included in the {\bf 16} representation of
SO(10), and there are no gauge-invariant {\bf 16}$^2$ or {\bf 16}$^3$
couplings in SO(10). In principle, one could imagine using either a {\bf
54} or {\bf 210} representation which  do allow both quadratic and cubic
couplings in the superpotential. Indeed, it might seem natural to
utilize one of these fields, which are often present as Higgs fields
used to break SO(10) down to some intermediate gauge group. An
interesting possibility utilizing the {\bf 210} was considered in
Ref.~\cite{GM}, where different possible directions within the {\bf 210}
were considered as inflaton candidates. There are however, two major
hurdles in this approach. The first is that the mass scale $\mu$ for the
Higgs field would typically be of order the GUT scale rather than $\sim
10^{-5}$ needed for inflation. Secondly, Starobinsky-type inflation
drives the field toward zero vacuum expectation value (vev), which in
this case would correspond to SO(10) symmetry restoration. Then one is
left with the problem (reminiscent of early problems associated with
degenerate vacua in supersymmetric GUTs \cite{supercosm}) of breaking
SO(10) after inflation, whereas normally it is assumed that the
appropriate choice of vacuum is determined during inflation. Finally, we
note that reheating is so efficient in a model with the inflaton
associated with a GUT-scale Higgs field that the reheating temperature
is very high, leading to the overproduction of gravitinos
\cite{bbb}.   

We are therefore led to consider a construction with an SO(10)-singlet
inflaton field. While there is no problem writing a superpotential as in
Eq.~\eqref{WZW} for a singlet, one must couple it to matter for
reheating in such a way as to preserve its inflationary evolution
and respect the other phenomenological constraints.
In the model discussed below, we will see that the fermionic partner of
the inflaton mixes with the neutrino sector, leading to a double-seesaw
structure, and the twin requirements of Planck-compatible inflation and
an acceptable reheating temperature place constraints on the parameters
of the neutrino mass matrix whose consistency with experimental data we
discuss. 

\subsubsection{Model}

We consider this scenario within an SO(10) model of grand unification
that breaks to an intermediate-scale gauge group $G_{\rm int}$ via a vev
of a  $\mathbf{210}$ representation at the GUT scale $M_{\rm GUT}$. The
intermediate-scale gauge group is subsequently broken to the Standard
Model (SM) group $G_{\rm SM} = {\rm SU}(3)_C \otimes {\rm SU}(2)_L
\otimes {\rm U}(1)_Y$ by vevs of a pair of $\mathbf{16}$ and
$\overline{\bf 16}$ representations at the intermediate scale $M_{\rm
int}$, and to ${\rm SU}(3)_C \otimes {\rm U}(1)_{\rm EM}$ symmetry via
vevs of the minimal supersymmetric Standard Model (MSSM) Higgs fields
$H_u$ and $H_d$ as usual. The MSSM Higgs fields are given by mixtures of
the $\mathbf{10}$, $\mathbf{16}$, and $\overline{\bf 16}$ fields as we
will see below. As a result, the symmetry-breaking chain we consider is
given by 
\begin{equation}
 {\rm SO}(10) \xrightarrow[{\bf 210}]{} 
 G_{\rm int} \xrightarrow[{\bf 16}, \overline{\bf 16}]{}
 G_{\rm SM} \xrightarrow[H_u, H_d]{}
 {\rm SU}(3)_C \otimes {\rm U}(1)_{\rm EM} ~.
\label{eq:symbrchain}
\end{equation}
The intermediate gauge symmetry $G_{\rm int}$ we obtain after the SO(10)
symmetry breaking depends on the vev of the ${\bf 210}$. 
We also consider the case where $M_{\rm GUT} = M_{\rm int}$,
namely, where the SO(10) gauge symmetry is broken into the Standard Model gauge
symmetry directly.   
We use the following notations for the SO(10) fields:
$\Sigma$ is the {\bf 210} representation that breaks SO(10) at the GUT scale,
$\Phi$ and $\bar{\Phi}$ are the {\bf 16} and $\overline{\bf 16}$
representations that break the theory to the MSSM, respectively,
$H$ is the {\bf 10} representation whose SU(2)$_L$ doublet components
mix with $\Phi$ and $\bar{\Phi}$ to yield the MSSM Higgs fields $H_u$
and $H_d$, $\psi_i$ ($i=1,2,3$) are the MSSM matter {\bf 16} multiplets
with $i$ the generation index, and $S_i$ $(i=1,2,3)$ denote the SO(10)
singlet {\bf 1} superfields, where one of these fields will be
identified as inflaton. The $R$-parity of each field is defined as
usual: $R \equiv (-1)^{3(B-L) + 2s}$ \cite{Farrar:1978xj}, where $B$,
$L$, and $s$ denote the baryon number, lepton number, and spin of the
field, respectively. Since the $B-L$ symmetry is a subgroup of SO(10),
the $R$-parity of each SO(10) representation is uniquely determined. 

The field content is similar to the SO(10) GUT in Ref.~\cite{GN}, which
uses a $\mathbf{16}$ rather than the more common $\mathbf{126}$ to break
the intermediate scale \cite{Clark:1982ai, bm, Aulakh:2002zr,
Aulakh:2003kg, Bajc:2004xe}. A supersymmetric version of this
``minimal'' theory was discussed in Ref.~\cite{Sarkar:2004ww}. In this
version of SO(10), the $\mathbf{126}$ and $\mathbf{\overline{126}}$ are
replaced by a pair of $\mathbf{16}$ and $\mathbf{\overline{16}}$, and
there is one singlet per generation, one of which is identified as the
inflaton. Since the $\psi \psi \Phi$ and $\psi \psi \bar{\Phi}$
couplings are forbidden by gauge symmetry, the vevs of the $\mathbf{16}$ and
$\mathbf{\overline{16}}$ fields do not generate Majorana mass terms for
right-handed neutrinos via renormalizable couplings. However, in our
model, non-zero light neutrino masses are induced via the mixing of
${\bf 1}$ and ${\bf 16}$ fields \cite{GN}, as we see in
Sec.~\ref{sec:neutrinomass}. In principle, only one such singlet is
needed for inflation, whereas two are needed for leptogenesis and the
non-zero neutrino mass differences, and three for non-zero neutrino
masses for all three neutrinos.

We consider the following generic form for the superpotential of the
theory:\footnote{To obtain the Starobinsky inflationary potential, we
drop a possible term linear in the singlet field $S$.}
\begin{align} \notag
W &= \frac{m}{2}S^2 - \frac{\lambda}{3} S^3 + y H\psi\psi + (M+bS)\bar{\Phi}\psi + m_{\Phi}\bar{\Phi}\Phi + \frac{\eta}{4!}\bar{\Phi}\Phi\Sigma + \frac{m_{\Sigma}}{4!}\Sigma^2 + \frac{\Lambda}{4!}\Sigma^3 \\[5pt] 
&+ m_{H}H^2 + \lambda_{SH} SH^2
+H(\alpha\Phi\Phi+\bar{\alpha}\bar{\Phi}\bar{\Phi}
+ \alpha^\prime \Phi \psi
) + cS\bar{\Phi}\Phi + \frac{b^\prime}{4!}\bar{\Phi}\psi\Sigma + \frac{\gamma}{4!}S\Sigma^2 + \kappa \,  \label{Wgen},
\end{align}
where for simplicity we have omitted the tensor structure of each term
and suppressed the generation indices. We assume that there is no mixing
among the singlet superfields $S_i$. 
The first two terms are the $S$-dependent Wess--Zumino superpotential
terms that reproduce the predictions of Starobinsky inflation in
no-scale supergravity~\cite{ENO6,ENO7}. The third term determines the SM
Yukawa couplings. The fourth and tenth terms include couplings between the
inflaton $S$ and SM fields: the magnitude of these couplings determines
the neutrino masses and the decay rate of the inflaton. The SM singlet
components of $\Phi$, $\bar{\Phi}$, and $\Sigma$ can acquire
non-vanishing vevs through the couplings included in the fifth through
eighth terms. After these fields develop vevs, the $\alpha H\Phi \Phi$
and $\bar{\alpha} \bar{H} \bar{\Phi} \bar{\Phi}$ terms induce mixing among the
SU(2)$_L$ doublet components inside $H$, $\Phi$, and $\bar{\Phi}$, and
by appropriately choosing these couplings we can make two linear
combinations of these fields, denoted by $H_u$ and $H_d$, much lighter
than the GUT and intermediate scales \cite{Sarkar:2004ww}, thereby realizing the
desirable doublet-triplet splitting. The vevs of these fields then break the SM
gauge group at the electroweak symmetry breaking scale as in the MSSM. 
In addition, after $\Phi$ acquires a vev, the $\alpha^\prime H\Phi \psi$
term induces an $R$-parity-violating term $H_u L$, where $L$ is the
SU(2)$_L$ doublet lepton field. This is because $\Phi$ is odd under
$R$-parity and thus its vev spontaneously breaks $R$-parity. On the
other hand, the other $R$-parity-violating operators in the MSSM, {\it i.e.},
$LL\bar{e}$, $LQ\bar{d}$, and $\bar{u} \bar{d} \bar{d}$, where
$\bar{e}$, $Q$, $\bar{u}$, and $\bar{d}$ are the SU(2)$_L$ singlet
charged lepton, doublet quark, singlet up-type quark, and singlet
down-type quark fields, respectively, are not generated at
renormalizable level. 
The constant $\kappa$ is tuned to yield a weak-scale gravitino mass
through the relation $m_{3/2}=\langle e^{K/2}W\rangle$: it may be
generated by the presence of a separate supersymmetry-breaking sector
such as a Polonyi sector \cite{pol}. The SO(10) no-scale \kahler
potential is then taken to be \cite{EKNN}
\beq\label{Kgen}
K=-3\ln\left[T+T^*-\frac{1}{3}\left(S^*S + H^{\dagger}H +
\psi^{\dagger}\psi + \Phi^{\dagger}\Phi + \bar{\Phi}^{\dagger}\bar{\Phi}
+ \frac{1}{4!}\Sigma^{\dagger}\Sigma \right)\right]\,, 
\eeq
which mimics the two-field prototype (\ref{no-scaleL}).

The superpotential in Eq.~\eqref{Wgen} contains several terms that are
additional to those in the minimal model in Ref.~\cite{Sarkar:2004ww},
notably those with the couplings $\lambda$, $\lambda_{SH}$, $M$,
$\alpha^\prime$, $c$, $b^\prime$, and $\gamma$. In
Ref.~\cite{Sarkar:2004ww}, there is an extra $\mathbb{Z}_2$ symmetry
(besides the SO(10) gauge symmetry) that forbids these couplings, which
is obtained by modifying the definition of $R$-parity as $R =
(-1)^{3(B-L) + 2s + \chi}$, where $\chi$ is a new quantum number: $1$
for $S$, $\Phi$, and $\bar{\Phi}$, and $0$ for the other
fields. However, for Starobinsky-type inflation, we must have a term
that is cubic in the singlet inflaton, thus we do not introduce such an
extra $\mathbb{Z}_2$ symmetry. Thus we are, in principle, allowed (even
obliged) to write down the additional couplings in Eq. (\ref{Wgen}). In
this case, $R$-parity is spontaneously broken when $\Phi$ and
$\bar{\Phi}$ develop vevs, as we mentioned above. For the most part, we
will assume these terms to be absent or small, but will comment on their
possible effects on our results. This assumption is stable against
radiative corrections thanks to the non-renormalization property of the
superpotential terms. We also comment in the following discussion on the
effect of $R$-parity violation in this theory.  

\subsubsection{Vacuum conditions}

The SO(10) and intermediate gauge symmetries are spontaneously broken by
SM singlet components of the above fields without breaking the SM gauge
group. Such components are contained in $\Sigma$, $\Phi$, $\bar{\Phi}$,
and $\psi$ (as well as $S$), and we denote these vevs by
\begin{align}
 p &= \langle \Sigma ({\bf 1}, {\bf 1}, {\bf 1}) \rangle ~, ~~~~~~
 a = \langle \Sigma ({\bf 15}, {\bf 1}, {\bf 1}) \rangle ~, ~~~~~~
 \omega= \langle \Sigma ({\bf 15}, {\bf 1}, {\bf 3}) \rangle ~,
 \nonumber \\
 \phi_R &= \langle \Phi (\overline{\bf 4}, {\bf 1}, {\bf 2}) \rangle ~,
~~~~~~
 \bar{\phi}_R = \langle \bar{\Phi} ({\bf 4}, {\bf 1}, {\bf 2}) \rangle ~,
~~~~~~
 \widetilde{\nu}_R = \langle \psi (\overline{\bf 4}, {\bf 1}, {\bf 2})
 \rangle ~, 
\label{eq:vevs}
\end{align}
where we express the component fields in terms of the ${\rm SU}(4)_C
\otimes {\rm SU}(2)_L \otimes {\rm SU}(2)_R$ quantum numbers. 
We assume that all of the vevs of $S_i$ vanish after
inflation; one of them which is regarded as inflaton is automatically
driven into zero after inflation as we see in the next section, while the
other two can also be stabilized at the origin by the quadratic 
coupling $m$. In addition, we will consider the cases where
$\widetilde{\nu}_R = 0$; otherwise, a non-zero vev of
$\widetilde{\nu}_R$ gives rise to a large $R$-parity violating term $H_u
L$ via the Yukawa coupling $y H\psi \psi$. We will see below that the
$\widetilde{\nu}_R = 0$ minimum is in fact stable with a positive
mass-squared if either $b$ or $b^\prime$ is non-zero. 
Depending on the values of $p$, $a$, $\omega$, $\phi_R$, and
$\bar{\phi}_R$, we obtain different symmetry-breaking patterns.
If all of these values are of the same order, then the SO(10) gauge
group is broken directly into the SM gauge group at the GUT scale. On
the other hand, if $\phi_R, \bar{\phi}_R \ll p, a,\omega$,
SO(10) is first broken into an intermediate gauge symmetry
$G_{\rm int}$ by vevs of $p$, $a$, and $\omega$ at the GUT scale, and
it is then subsequently broken by $\phi_R$ and $\bar{\phi}_R$
into $G_{\rm SM}$, as shown in Eq.~\eqref{eq:symbrchain}. 
We will discuss possible values of these vevs as well as the
corresponding intermediate gauge symmetries in what follows.

In the supersymmetric limit, all of the other components have
vanishing vevs. After the supersymmetry-breaking effects are
transmitted to the visible sector, certain linear combinations of the
doublet components $H$, $\Phi$, and $\bar{\Phi}$ develop vevs of the
order of the electroweak scale to break the SM gauge
symmetry spontaneously to ${\rm SU}(3) \otimes {\rm U}(1)_{\rm EM}$,
just as in the MSSM. The rest of the components in $S$, $H$, $\Phi$,
$\bar{\Phi}$, and $\Sigma$ are stabilized at the origin with GUT- or
intermediate-scale masses.

 In no-scale supergravity with a $T$-independent superpotential, the
 $F$-term part of the scalar potential has the simple form
 \cite{EKNN,EKN3} 
\beq
V=e^{2K/3}|W^i|^2\,.
\eeq
To study the scalar potential, we write the superpotential \eqref{Wgen}
 in terms of the SM singlet fields, with the rest of the fields set to
zero:  
\begin{align}\notag
W &=  \frac{m}{2}S^2 
- \frac{\lambda}{3} S^3 
 - (M+ b S)\bar{\phi}_R\nu_R 
+ (\eta\phi_R+ b^\prime\nu_R) \bar{\phi}_R(p+3a+6\omega) \\
&-(m_{\Phi} + cS) \bar{\phi}_R\phi_R
+ (m_{\Sigma}+\gamma S)(p^2+3a^2+6\omega^2) +2\Lambda(a^3+3p\omega^2+6a\omega^2) + \kappa ~.
\label{superpot}
\end{align}
As we discussed above, we study vacua where $S = \widetilde{\nu}_R =
0$. We also require that the non-zero vevs of $p$, $a$, $\omega$,
$\phi_R$, and $\bar{\phi}_R$ do not break supersymmetry. Therefore, the
$F$-terms of these fields should vanish, leading to the following set of
algebraic equations: 
\begin{align}
 2m_{\Sigma}p + 6\Lambda \omega^2 + \eta \phi_R\bar{\phi}_R &=0\,,
 \label{eq:vevp}\\ 
2m_{\Sigma}a + 2\Lambda (a^2+2\omega^2) + \eta \phi_R\bar{\phi}_R &=0\,,
\label{eq:veva}\\
2m_{\Sigma}\omega + 2\Lambda(p+2a) \omega + \eta \phi_R\bar{\phi}_R
 &=0 \label{eq:vevw}\,, \\
\bar{\phi}_R \left[-m_{\Phi}+\eta(p+3a+6\omega)\right] &= 0\,, \label{eq:vevphi}\\
\phi_R \left[-m_{\Phi}+\eta(p+3a+6\omega)\right] &= 0\,, \label{eq:vevphibar}\\
-c\phi_R\bar{\phi}_R + \gamma(p^2+3a^2+6\omega^2) &=0 \label{eq:vevsing}\,,\\ 
\bar{\phi}_R\left[-M+b^\prime(p+3a+6\omega)\right] &= 0\label{eq:vevnur}\,,
\end{align}
for $p$, $a$, $\omega$, $\phi_R$, $\bar{\phi}_R$, $S$, and
$\widetilde{\nu}_R$, respectively.
As discussed in Refs.~\cite{Bajc:2004xe,Aulakh:2003kg}, these equations possess
a variety of solutions that lead to different, degenerate
symmetry-breaking vacua, along with the SO(10)-preserving vacuum
$p=a=\omega=\phi_R=\bar{\phi}_R=0$. 
This solution can be parametrized in the form
\begin{alignat}{2} 
& p= -\frac{m_{\Sigma}}{\Lambda}\,\frac{x(1-5x^2)}{(1-x)^2}\,,
 \qquad\quad &&\quad\ \ \,
 a=-\frac{m_{\Sigma}}{\Lambda}\,\frac{1-2x-x^2}{1-x}\,, \nonumber\\
& \omega= \frac{m_{\Sigma}}{\Lambda}x\,,   &&\phi_R\bar{\phi}_R=
 \frac{2m_{\Sigma}^2}{\eta\Lambda}\,\frac{x(1-3x)(1+x^2)}{(1-x^2)} 
\label{omphi} \,,
\end{alignat}
where the parameter $x$ is a solution of the cubic equation
\beq\label{xcubic}
8x^3-15x^2+14x-3 = (x-1)^2\frac{\Lambda m_{\Phi}}{\eta m_{\Sigma}}\,,
\eeq
and where $|\phi_R|=|\bar{\phi}_R|\equiv \phi$ in order to ensure the
vanishing of the $D$-terms. This solution is identical to that found in
SO(10) models using a $\mathbf{126}$ \cite{Bajc:2004xe,Aulakh:2003kg}
rather than a $\mathbf{16}$, with the change in sign for $\omega$ in
Eq.~\eqref{omphi} and a sign change in the right-hand side of
Eq.~\eqref{xcubic}. The solutions in Eq.~\eqref{omphi} satisfy
Eqs.~\eqref{eq:vevp}--\eqref{eq:vevphibar}.
The conditions
\eqref{eq:vevsing} and \eqref{eq:vevnur} then restrict the parameters
$c$, $\gamma$, $M$, and $b^\prime$. 
From these equations we find that $\phi \ll p, a,
\omega$ is realized when $x \simeq 0$, 1/3, or $\pm i$
\cite{Bajc:2004xe}. For $x \simeq 
0$, Eq.~\eqref{omphi} is satisfied for $\Lambda m_\Phi/\eta m_\Sigma
\simeq -3$. In this case, $p \simeq \omega \simeq \phi \simeq 0$ and $a
= -m_\Sigma /\Lambda$, and we obtain $G_{\rm int} = {\rm SU}(3)_C
\otimes {\rm SU}(2)_L \otimes {\rm SU}(2)_R \otimes {\rm U}(1)_{B-L}
\otimes D$, where $D$ denotes $D$-parity \cite{Kuzmin:1980yp}. For $x
\simeq 1/3$, we need $\Lambda m_\Phi/\eta m_\Sigma \simeq 2/3$, which
leads to $p \simeq a \simeq -\omega \simeq -m_\Sigma/3\Lambda$ and
$G_{\rm int} = {\rm SU}(5) \otimes {\rm U}(1)$. In this case, the vevs
of $\phi_R$ and $\bar{\phi}_R$ cannot break SU(5), and thus the
intermediate gauge symmetry is broken by the difference among $p$, $a$,
and $-\omega$, whose sizes are  ${\cal O}(M_{\rm int})$; {\it i.e.},
$p-a$, $p+\omega ={\cal O}(M_{\rm int})$. This intermediate gauge
symmetry looks phenomenologically implausible, however, since the SU(5)
gauge bosons whose masses are ${\cal O}(M_{\rm int})$ cause rapid proton
decay. The $x \simeq \pm i$ case
is realized for $\Lambda m_\Phi/\eta m_\Sigma \simeq -3\pm 6i$, where we
obtain $p\simeq 3 m_\Sigma/\Lambda$, $a \simeq -2 m_\Sigma/\Lambda$,
$\omega \simeq \pm i m_\Sigma/\Lambda$, and $G_{\rm int } = {\rm
SU}(3)_C \otimes {\rm SU}(2)_L \otimes {\rm U}(1)_R \otimes {\rm
U}(1)_{B-L}$. Among the possibilities which lead to $\phi \ll p, a,
\omega$, we consider here only the case with $x \simeq 0$. 

In the above analysis we have assumed that $\widetilde{\nu}_R = 0$ at
the vacua. To check that this is indeed the case, we consider the scalar
potential terms that contain $\widetilde{\nu}_R$: 
\begin{align}
 \left|\frac{\partial W}{\partial S}\right|^2
&+ \left|\frac{\partial W}{\partial \bar{\phi}_R} \right|^2
+ \left|\frac{\partial W}{\partial p}\right|^2
+ \left|\frac{\partial W}{\partial a}\right|^2
+ \left|\frac{\partial W}{\partial \omega}\right|^2
\rightarrow \left[
|b|^2 + 46 |b^\prime|^2 
\right]\left|\widetilde{\nu}_R\right|^2 ~,
\end{align}
where we have used the conditions \eqref{eq:vevp}--\eqref{eq:vevnur}.
We see immediately that $\widetilde{\nu}_R$ has a positive mass term unless $b =
b^\prime = 0$, and thus is indeed stabilized at the origin in the vacua considered
above.  

Generically, the non-zero vevs of these fields lead to an ${\cal
O}(M_{\rm GUT})$ contribution to supersymmetry breaking, which may be
fine-tuned with $\kappa$ of ${\cal O}(M_{\rm GUT})$ to be of the order
of supersymmetry-breaking scale, $M_{\rm SUSY}$. We note, however, that one
particular solution is obtained if we further require that the GUT
sector does not require a fine-tuning  of $\kappa$ to ensure a
weak-scale gravitino mass, {\it i.e.}, if we impose $\kappa\ll M_{\rm
GUT}$. This minimum with vanishing superpotential is found for \cite{GM}
$x \simeq -0.3471$ and $\Lambda m_{\Phi}/\eta m_{\Sigma} \simeq -5.5115$,
in which case we have $(p,a,\omega) \simeq (-0.0138, 0.2120, 0.0630)
m_{\Phi}/\eta$ with $\phi\simeq 0.3985 \sqrt{m_{\Phi}  m_{\Sigma}}/\eta$
in units of $M_P$. However, we do not consider this particular solution
here.

\subsubsection{Doublet-triplet splitting}
\label{sec:dtspl}

After the above fields acquire vevs, we obtain the MSSM as an effective
theory. To realize electroweak symmetry breaking correctly, we need
the $\mu$-term in the MSSM to be of the order of the
supersymmetry-breaking scale, which is assumed to be much lower than the
GUT scale. This is the so-called doublet-triplet splitting, and in this
model we can realize this by fine-tuning the $\alpha$ and $\bar{\alpha}$
couplings in Eq.~\eqref{Wgen}. To see this, let us first write down the
relevant superpotential terms:
\begin{align}
 W \ni m_H H_L \bar{H}_L - \alpha {H}_L \phi_L \phi_R 
- \bar{\alpha} \bar{H}_L \bar{\phi}_L \bar{\phi}_R + \left[m_\Phi
+ \eta (p-3a)
\right] \bar{\phi}_L \phi_L ~,
\label{eq:wdoub}
\end{align}
where $H_L$ and $\bar{H}_L$ are the SU(2)$_L$ doublet components of $H$
with $Y=+1/2$ and $-1/2$, respectively, and $\phi_L$ ($\bar{\phi}_L$) is
the SU(2)$_L$ doublet component in $\Phi$ ($\bar{\Phi}$) with $Y=-1/2$
($1/2$). After $\phi_R$ and $\bar{\phi}_R$ develop a vev, $\phi$, 
Eq.~\eqref{eq:wdoub} leads to 
\begin{equation}
 W_{\mu} = (\bar{H}_L, {\phi}_L) 
\begin{pmatrix}
 m_H & -\bar{\alpha} \phi \\ -{\alpha} \phi & m_\Phi + \eta(p-3a)
\end{pmatrix}
\begin{pmatrix}
 H_L \\ \bar{\phi}_L
\end{pmatrix}
~.
\label{eq:wmu}
\end{equation}
We note that $m_\Phi + \eta(p-3a) = 2\eta (p + 3\omega)$ when the conditions
\eqref{eq:vevphi} and \eqref{eq:vevphibar} are applied. The mass matrix
in Eq.~\eqref{eq:wmu} may be diagonalized using two unitary matrices
${\cal U}$ and ${\cal D}$:
\begin{equation}
 {\cal D}^T 
\begin{pmatrix}
 m_H & -\bar{\alpha} \phi \\ -{\alpha} \phi & 2 \eta(p+3\omega)
\end{pmatrix}
{\cal U}
= 
\begin{pmatrix}
 \mu_1 & 0 \\ 0& \mu_2
\end{pmatrix}
~,
\end{equation}
with 
\begin{equation}
 \mu_{1,2} = \frac{1}{2} \left[
m_H + 2 \eta(p+3 \omega) \mp \sqrt{
\left[m_H + 2 \eta(p+3\omega)\right]^2 -4\Delta
}
\right]~,
\end{equation}
and
\begin{equation}
 \Delta = 2 \eta m_H (p+3\omega) - \alpha
  \bar{\alpha}\phi^2 ~.
\label{eq:deltadef}
\end{equation}
Thus, to obtain a $\mu$-term of order supersymmetry-breaking scale, we
need to fine-tune $\Delta$ to be much smaller than ${\cal O}(M_{\rm
GUT}^2)$. This can be realized by cancelling the first and second terms
in Eq.~\eqref{eq:deltadef}. If $\phi ={\cal O}(M_{\rm GUT})$, $p$,
$\omega$, and $m_H$ can also be ${\cal O}(M_{\rm GUT})$ to achieve the
fine-tuning. If $\phi \simeq M_{\rm int} \ll M_{\rm GUT}$, on the other
hand, we need $\eta m_H \simeq M_{\rm int}^2/M_{\rm GUT}$ unless $p\simeq -3\omega$, {\it i.e.}, $x \simeq
(3\pm i \sqrt{7})/8$. In this case, $m_H$ and/or $m_\Phi$ are much
smaller than the GUT scale (notice that there is a relation between
$m_\Phi$ and $\eta$ via the conditions \eqref{eq:vevphi} and
\eqref{eq:vevphibar}), which may be phenomenologically dangerous as we
discuss below.  For this reason, we will concentrate on models in which the
intermediate scale is close to the GUT scale, which as we will see is
beneficial for proton decay and the evolution the Higgs fields during
inflation. This case is also favored in terms of gauge coupling
unification, as the gauge couplings in the MSSM meet each other around
$2\times 10^{16}$~GeV with great accuracy, which implies the absence of
an intermediate scale below the GUT scale. In any case, all of the
components in the mass matrix should be of the same order in order for the
cancellation to occur, and thus the mixing angles in ${\cal U}$ and
${\cal D}$ are ${\cal O}(1)$.

The eigenstates of the matrix in Eq.~\eqref{eq:wmu} are related to the
doublet fields via 
\begin{equation}
\begin{pmatrix}
 H_u \\ H_u^\prime
\end{pmatrix}
=
{\cal U}^\dagger 
 \begin{pmatrix}
  H_L \\ \bar{\phi}_L
 \end{pmatrix}
~,~~~~~~
\begin{pmatrix}
 H_d \\ H_d^\prime
\end{pmatrix}
=
{\cal D}^\dagger 
 \begin{pmatrix}
  \bar{H}_L \\ {\phi}_L
 \end{pmatrix}
~,
\label{eq:udmixing}
\end{equation}
where $H_u$ and $H_d$ are to be regarded as the MSSM Higgs fields with a
$\mu$-term of $\mu_1 \ll {\cal O}(M_{\rm GUT})$, while the heavier
states $H_u^\prime$ and $H_d^\prime$ have 
\begin{equation}
 \mu_2 \simeq m_H + 2\eta (p+3\omega) ~.
\end{equation}
After supersymmetry is broken, $H_u$ and $H_d$ develop vevs to break
electroweak symmetry, while $H_u^\prime$ and $H_d^\prime$ remain at
the origin.

Finally we note that the fine tuning discussed above could potentially
be avoided if instead of SO(10), the GUT gauge group were flipped
${\rm SU}(5) \otimes {\rm U}(1)$ \cite{flipped1}.  In this case the
doublet-triplet separation is solved by a missing partner mechanism
\cite{flipped2}. As we will note below, several of the wanted features
discussed below could be carried over to a flipped model, though we do
not work out such a model in any detail here.

\subsubsection{Proton decay}

The $\alpha$ and $\bar{\alpha}$ couplings in Eq.~\eqref{Wgen} also
induce mixing between the color-triplet components in $H$ with those
in $\Phi$ and $\bar{\Phi}$. Due to  the ${\bf 210}$ vevs, the
vector-like mass term for the color-triplets in $\Phi$ and $\bar{\Phi}$
is different from that for $\phi_L$ and $\bar{\phi}_L$. Therefore, even
though we have fine-tuned $\Delta$ to obtain $\mu_1$ of ${\cal O}(M_{\rm
SUSY})$, this does not result in an ${\cal O}(M_{\rm SUSY})$ $\mu$-term
for the color-triplet multiplets. In particular, for ${\cal O}(M_{\rm
GUT})$ values of $m_H$, $m_\Phi$, $\phi$, we have ${\cal O}(M_{\rm
GUT})$ $\mu$-terms for the color-triplet components. On the other hand,
if (some of) these values are much smaller than ${\cal O}(M_{\rm GUT})$,
then the color-triplet Higgs masses may also be much lighter than the
GUT scale.

The exchange of the color-triplet Higgs multiplets leads to proton
decay, {\it e.g.}, via $p\to K^+ \bar{\nu}$, and in many supersymmetric GUTs
this turns out to be the dominant contribution \cite{Sakai:1981pk}. 
If supersymmetry breaking is TeV-scale, the resultant proton decay lifetime
tends to be too short \cite{Goto:1998qg, mp}, and thus some additional
mechanism is required to suppress this contribution. A simple way to
evade the proton decay bound is to take $M_{\rm SUSY}$ in the
multi-TeV region \cite{Hisano:2013exa, Ellis:2016tjc}; for instance, in
the CMSSM, the current limit $\tau_p(p\to K^+ \bar{\nu}) > 6.6 \times
10^{33}$~yrs \cite{Takhistov:2016eqm, Abe:2014mwa} is satisfied for $m_0
= 10$~TeV, $m_{1/2} = 8$~TeV, $A_0 = 0$, $\tan \beta = 5$
\cite{Ellis:2016tjc}. In the present scenario, however, the proton decay bound may become more severe. First, in SO(10)
GUT models, a large $\tan\beta \simeq m_t/m_b$ is favored to realize the
SO(10) relation for the Yukawa couplings. Since the wino (higgsino)
exchange contribution to the $p\to K^+ \bar{\nu}$ decay amplitude is
proportional to $1/\sin 2\beta \simeq (\tan \beta)/2 $, such a large $\tan \beta$ enhances the proton
decay rate by orders of magnitude. Secondly, as we see above, if $M_{\rm
int} \simeq \phi \ll M_{\rm GUT}$, the color-triplet Higgs masses tend
to be as light as the intermediate scale. Since the proton decay rate
is inversely proportional to the square of the color-triplet Higgs mass, this
again reduces the proton lifetime by orders of magnitude. We do not discuss these issues further in this paper,
simply assuming that the proton decay limit is evaded because of a very
high supersymmetry-breaking scale and/or some additional mechanism to
suppress the color-triplet Higgs exchange contribution. As noted earlier, 
these issues are automatically solved in a flipped SU(5) model, but here we 
will concentrate on models in which the intermediate scale is close to or at the GUT scale
to minimize the latter effect on proton decay.

Of course, the exchange of the GUT-scale gauge bosons also induces
proton decay, where $p \to e^+ \pi^0$ is the dominant decay channel. The
lifetime of the decay channel is approximated by
\begin{equation}
 \tau (p \to e^+ \pi^0) \simeq 
5\times 10^{34} \times
\left(\frac{1/25}{\alpha_{\rm GUT}}\right)^4
\left(\frac{M_X}{10^{16}~{\rm
GeV}}\right)^4 
\left(\frac{3}{A_R}\right)^2
~{\rm years} ~,
\end{equation}
where $\alpha_{\rm GUT} = g_{\rm GUT}^2/(4\pi)$ is the unified gauge
coupling, $M_X$ denotes collectively the GUT-scale gauge boson masses,
and $A_R$ is a renormalization factor.\footnote{The one-loop renormalization
factors of the K\"{a}hler type proton-decay operators for each
intermediate gauge symmetry in supersymmetric theories are
given in Ref.~\cite{Munoz:1986kq}. For a two-loop-level computation, see
Ref.~\cite{Hisano:2013ege}. Below the supersymmetry-breaking scale,
renormalization factors are given at one-loop level in
Ref.~\cite{Abbott:1980zj}. Below the electroweak scale, we use the QCD
renormalization factors computed at two-loop level in
Ref.~\cite{Nihei:1994tx}. The relevant hadron matrix elements are
evaluated in Ref.~\cite{Aoki:2013yxa}.} 
The GUT-scale gauge boson masses can be expressed in terms of $p$, $a$,
$\omega$, and $\phi$ as well as the unified gauge coupling; for
instance, the $({\bf 3}, {\bf 2}, -\frac{5}{6})\oplus (\overline{\bf 3},
{\bf 2}, \frac{5}{6})$ components (in terms of the SM quantum numbers)
of the SO(10) gauge boson has a mass $g_{\rm GUT} \sqrt{4|a +
\omega|^2 +2|p+\omega|^2}$ \cite{Aulakh:2004hm}, which shows that the
current experimental bound $p\to e^+ \pi^0 >1.7\times 10^{34}$~yrs
\cite{Takhistov:2016eqm} is evaded if these vevs are $\gtrsim
10^{16}$~GeV.

\section{Realization of Inflation}
\label{inflation}

As was explained in the previous Section, in our model the singlet $S$
plays the role of the inflaton. The shape of its effective potential is
dependent on the couplings of $S$ to itself and to the Higgs
sector. Strictly speaking, the Starobinsky potential is realized via the
first two terms in Eq.~\eqref{superpot} whereas the other terms in the
superpotential involving $S$, namely those proportional to couplings $b,
c$, and $\gamma$, all break the scale symmetry associated with the
potential. Therefore in order to realize suitable inflation, we must
require these couplings to be small. For now, we take $c = \gamma = 0$
and comment later on the effects if they are non-zero, while noting that
$b$ should be non-zero as it also enters into the neutrino mass matrix,
as we discuss in the following Section. 

Sufficient inflation would require at least $ N_* \simeq 50$ $e$-folds
of expansion, where 
\beq
N_* = -\int_{s_*}^0 \frac{1}{\sqrt{2\epsilon}} \, : \qquad \epsilon =
\frac{1}{2}(V_s/V)^2 
\eeq
for a potential $V(s)$, where $s$ is the canonically normalized
inflaton. For the Starobinsky potential, a total number of $e$-folds $N
> N_* =  50 (60)$ is found for an initial value of $s$, $s_i > s_* =
5.24 (5.45)$. Thus, to realize Starobinsky-like inflation, we must
ensure that any significant deviation from the Starobinsky potential
occurs at values of $s > 5.24$.

During inflation, the GUT-breaking Higgs fields are displaced from their
vacuum values (\ref{omphi}). These displacements would be exponentially
small for $b=0$: the potential derivatives with respect to
$p,a,\omega,\phi$ all vanish in the limit $S\rightarrow \sqrt{3}$ if the
corresponding values of the Higgs singlet components are given by
(\ref{omphi}). For a finite, but large, value of the canonically
normalized inflaton $s$, defined along the real direction as 
\beq
S \equiv \sqrt{3}\tanh(s/\sqrt{6})\,,
\eeq
the instantaneous deviation from the vacuum vev is proportional to $m^2e^{-2\sqrt{2/3}\,s}$. 

For a non-vanishing but small value of $b$, the instantaneous minima of the singlets during inflation 
are perturbed relative to (\ref{omphi}) by an $\mathcal{O}(b^2)$ factor; for example,
\beq
\delta \phi \simeq b^2\,f(x;m_{\Sigma},m_{\Phi},\eta)\,.
\eeq
where $f$ is a (somewhat complicated and long) function of $x$ and the superpotential parameters. 
This function is divergent for $x=0,1/3$ and $\pm i$, the values that give rise naturally to the hierarchy 
$M_{\rm GUT}\gg M_{\rm int}$. We have checked numerically that, for $x$ sufficiently close to these singular points, 
any finite value of $b$ will drive $\phi$ to zero during and after inflation, 
preventing the spontaneous breaking of the intermediate gauge group.

\begin{figure}[!h]
\centering
    \scalebox{0.80}{\includegraphics{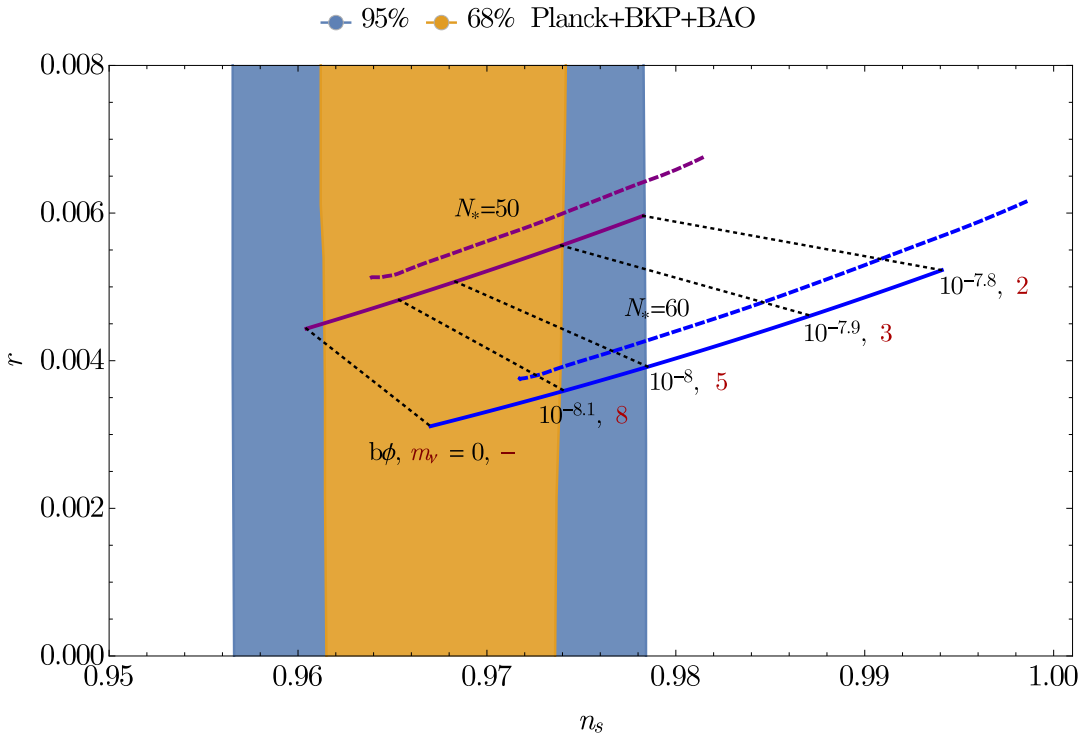}}
    \caption{\em Parametric $(n_s,r)$ curves as functions of $b\phi$ for $N_*=50,60$, with
    the 68 and 95\% CL Planck constraints shown in the background. The solid curves illustrate the parametric dependence using the analytical approximation (\ref{Vnotfull}) and (\ref{deltaV}) assuming $\Delta K = 0$. The dashed curves show the power spectrum parameters calculated numerically with $x=-1$, for the same range of $b\phi$.
The dotted curves illustrate particular values of $b\phi$, quantified in units of $M_P$, 
and we indicate the corresponding left-handed neutrino masses in units of $10^{-4}$ eV assuming 
$f_\nu \sin \beta = 10^{-5}$ and $m = 10^{-5} M_P$. See Section~\ref{sec:neutrinomass} for more details
of the relation between the light neutrino masses and $b \phi$. 
        }
    \label{fig:planck1}
\end{figure}

\begin{figure}[!h]
\centering
    \scalebox{0.80}{\includegraphics{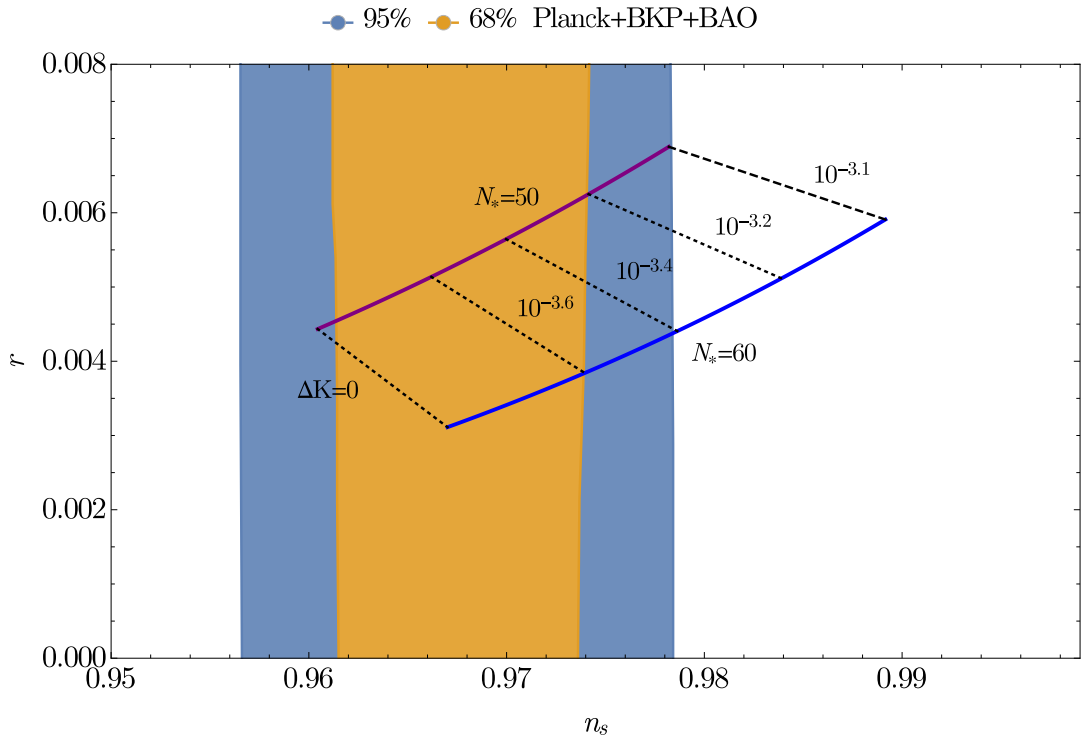}}
    \caption{\em As in Figure \protect\ref{fig:planck1}, but for different values of $\Delta K \equiv |p|^2 + 3|a|^2 + 6|\omega|^2 + 2|\phi|^2$, assuming $b\phi=0$.
        }
    \label{fig:planck2}
\end{figure}

Let us for now assume that the Higgs fields are displaced a negligible amount from their vacuum values during inflation, 
$\{p,a,\omega,\phi\}\simeq \{p_0,a_0,\omega_0,\phi_0\}$.
In this case, the scalar potential during inflation takes the simple form 
\beq \label{Vsimple}
V\simeq \frac{\hat{V}}{\left[1-\frac{1}{3}(|S|^2 + |p|^2 + 3|a|^2 + 6|\omega|^2 +2|\phi|^2  )\right]^2} \, ,
\eeq
where
\beq\label{Vbcg}
\hat{V}= |m S - \lambda S^2|^2+|S|^2\Big[|b \phi |^2+2|c\phi|^2+|2\gamma p|^2 + |6\gamma a|^2 + |12\gamma\omega|^2\Big]\,.
\eeq
This shows that, in order to recover the predictions of no-scale Starobinsky-like inflation, 
we need to constrain independently the values of the squared moduli inside the brackets.
For $c=\gamma=0$,
we find in terms of the canonically-normalized field $s$ that for $\lambda=m/\sqrt{3}$ the scalar potential takes the form
\begin{align}\notag
V &= \left(1-\tanh ^2(s/\sqrt{6})-\tfrac{1}{3}( |p|^2 + 3|a|^2 + 6|\omega|^2+2|\phi|^2) \right)^{-2} \\ \label{Vfull}
&\quad\qquad \times 3 \tanh ^2 (s/\sqrt{6} ) \left[ m^2\left(\tanh (s/\sqrt{6} ) -1\right)^2 + |b\phi|^2  \right]\\ \label{Vnotfull}
&\simeq \frac{3}{4}m^2\left(1-e^{-\sqrt{2/3}\,s}\right)^2 +\Delta V\,,
\end{align}
where
\beq\label{deltaV}
\Delta V = \left[\frac{3}{4} |b\phi|^2 + \frac{1}{2}m^2 e^{-\sqrt{2/3}\,s}\left(|p|^2 + 3|a|^2 + 6|\omega|^2 + 2|\phi|^2\right) \right]  \sinh ^2 (\sqrt{2/3}\, s )\,.
\eeq
We show in Figs. \ref{fig:planck1} and \ref{fig:planck2} the effects of the coupling $b$ and the 
quantity $\Delta K \equiv |p|^2 + 3|a|^2 + 6|\omega|^2 + 2|\phi|^2$ in $\Delta V$.
In each Figure, we plot the slope of the perturbation spectrum, $n_s$ and the tensor-to-scalar ratio, $r$
given by (the quantity $\eta \equiv V_{ss}/V$ here is not to be confused with the superpotential coupling):
\beq
n_s \simeq 1 - 6 \epsilon + 2 \eta \, , \qquad r \simeq 12 \epsilon  \, .
\eeq
The orange (purple) shaded regions correspond to the 68 (95) \% CL limits from 
Planck \cite{planck15}. 
In the limit where $b\phi,\Delta K \ll 1 $, the inflationary parameters can be approximated analytically by
\begin{align}\label{nsan}
n_s &\;\simeq\; -\frac{2}{N_*} + \frac{8}{3}\left(\frac{b\phi}{m}\right)^2N_*^2 + \frac{32}{81}\,\Delta K\,N_*\,,\\ \label{ran}
r &\;\simeq\; \frac{12}{N_*^2}  + \frac{32}{3}\left(\frac{b\phi}{m}\right)^2N_* + \frac{64}{27}\,\Delta K\,.
\end{align}

\begin{figure}[h!]
\centering
    \scalebox{0.57}{\includegraphics{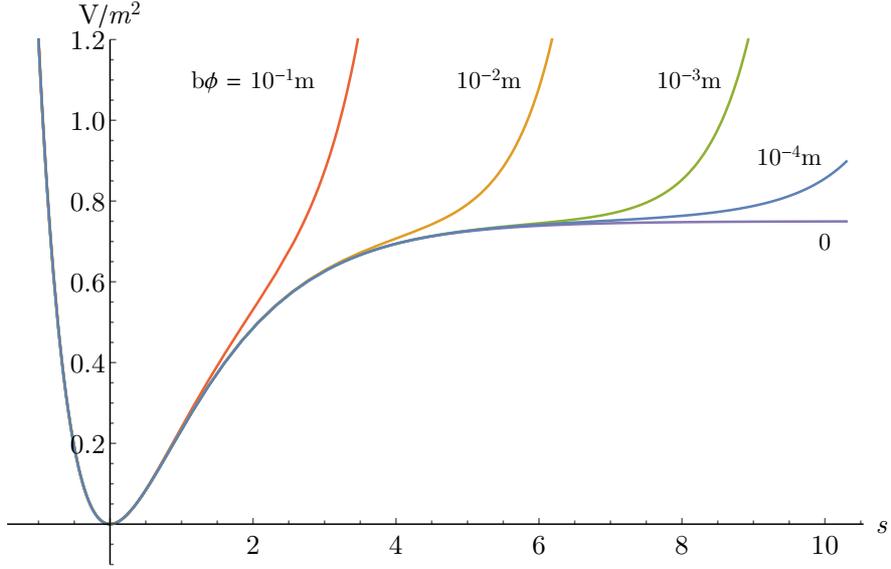}}
    \caption{\em The inflationary potential for different values of $b\phi$, in units of the inflaton mass $m\simeq 10^{-5}M_P$. The curve labeled $b\phi = 0$ is the Starobinsky potential. We assume $\Delta K = 0$ here.
        }
    \label{fig:Vcorr1}
\end{figure}

We see in Fig.~\ref{fig:planck1} the effect of a non-zero value of $b$.
The solid curves show the positions in the $(n_s,r)$ plane for $N_* =
50$ and 60~\footnote{The dotted lines simply interpolate between $N_* =
50$ and 60.} as $b\phi$ is increased from 0 to $10^{-7.8}$ using the
analytical approximation for the potential given by
Eqs.~\eqref{Vnotfull} and (\ref{deltaV}). Here we have taken $\Delta K
= 0$, and recall that $b\phi = 0$ corresponds to the exact Starobinsky
result. The dashed lines
are derived from a full numerical evolution. For these solutions, 
$\Delta K \approx 10^{-3.7}$ as would be
obtained for $x = -1$. This is the cause of the offset when $b \phi = 0$.
In order to obtain values of $(n_s, r)$ consistent with Planck, we must
require that the product $b \phi < 10^{-7.8}$ $(10^{-8})$ for $N_*\simeq
50$ $(60)$ $e$-folds of inflation. Since the vev of $\Phi$ is no larger
than the GUT scale, $\phi \la 10^{-2.3}$, the most severe constraint we
have on the coupling $b$ is $b < 10^{-5.7}$. The scalar potential for
several choices of $b\phi$ is shown in Fig. \ref{fig:Vcorr1}. As one can
see, so long as $b\phi \la 10^{-2.5} m \sim 10^{-7.5}$, the potential is
indistinguishable from the Starobinsky potential out to the value $s
\sim 5.5$ needed for 60 $e$-folds of inflation. 

\begin{figure}[t!]
\centering
    \scalebox{0.57}{\includegraphics{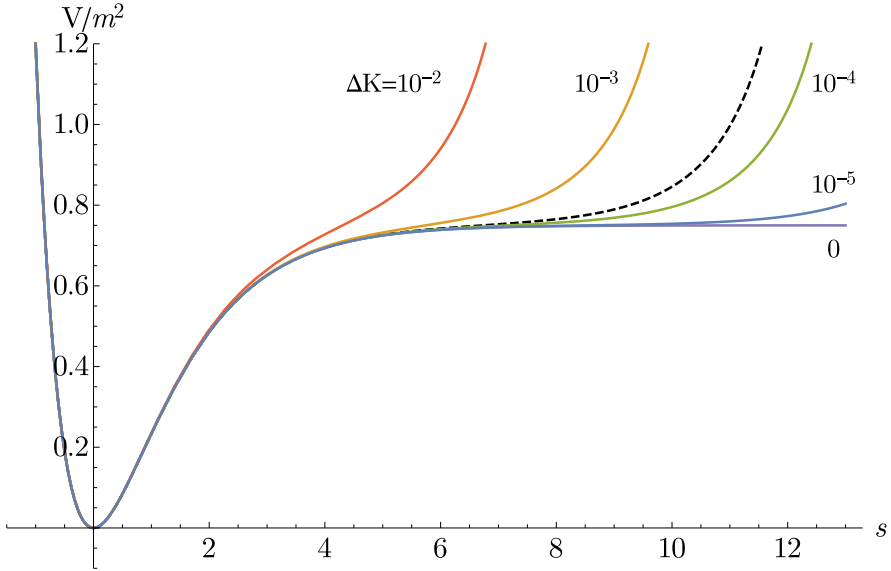}}
    \caption{\em The inflationary potential for different values of $\Delta K \equiv |p|^2 + 3|a|^2 + 6|\omega|^2 + 2|\phi|^2$, for $b=0$. The black dashed line corresponds to $\phi = p=a=\omega=10^{16}\,{\rm GeV}$.
        }
    \label{fig:Vcorr2}
\end{figure}

We see in Fig.~\ref{fig:planck2} the corresponding effect of varying
$\Delta K$ for $b\phi = 0$. As we have fixed the value of $\Delta K$, we
show here only the analytic result. In this case, in order to obtain
values of $(n_s, r)$ consistent with Planck, we must require that the
quantity $\Delta K  < 10^{-3.1}$ $(10^{-3.4})$ for $N_*\simeq 50$ $(60)$
$e$-folds of inflation. If the largest vevs associated with $p, a$
and/or $\omega$ are of order $10^{16}$~GeV, $\Delta K \la10^{-3.7}$ and
the bounds from Planck are always satisfied. The scalar potential for
this case is shown by the dashed curve in Fig.~\ref{fig:Vcorr2}, and is
Starobinsky-like out to $s \approx 8$.  Figure~\ref{fig:Vcorr2} also shows
the potential for other choices of $\Delta K$ for $b = 0$.  

For generic values of $b$ and $\Delta K$, we can  approximate numerically the limits
on $\Delta K$ and $b \phi$ by \beq\label{combinedKbphi}
\Delta K + (2\times 10^6\, b \phi )^2 \le \begin{cases}
0.00078, & N_* = 50 \, , \\ 
0.00043, & N_* = 60 \, . \end{cases} 
\eeq

So far we have relied on the assumption that the Higgs fields track the
instantaneous minimum during inflation. We have verified this behavior
by integrating numerically  the classical equations of motion, given by 
\beq\label{eomfull}
\ddot{\Psi}^a + 3H\dot{\Psi}^a + \Gamma^a_{bc}\dot{\Psi}^b\dot{\Psi}^c + K^{a\bar{b}}\frac{\partial V}{\partial \bar{\Psi}^{\bar{b}}} \;=\; 0\,.
\eeq
Here the indices run over all field components, with
$\Psi^a\equiv\{T,S,p,a,\omega,\phi_R,\cdots\}$, $K^{a\bar{b}}$ denotes
the inverse \kahler metric, and the connection coefficients are given by 
\beq
\Gamma^a_{bc} = K^{a\bar{d}}\partial_b K_{c\bar{d}}\,.
\eeq

\begin{figure}[!p]
\centering
        \hspace{0.3cm} \scalebox{0.68}{\includegraphics{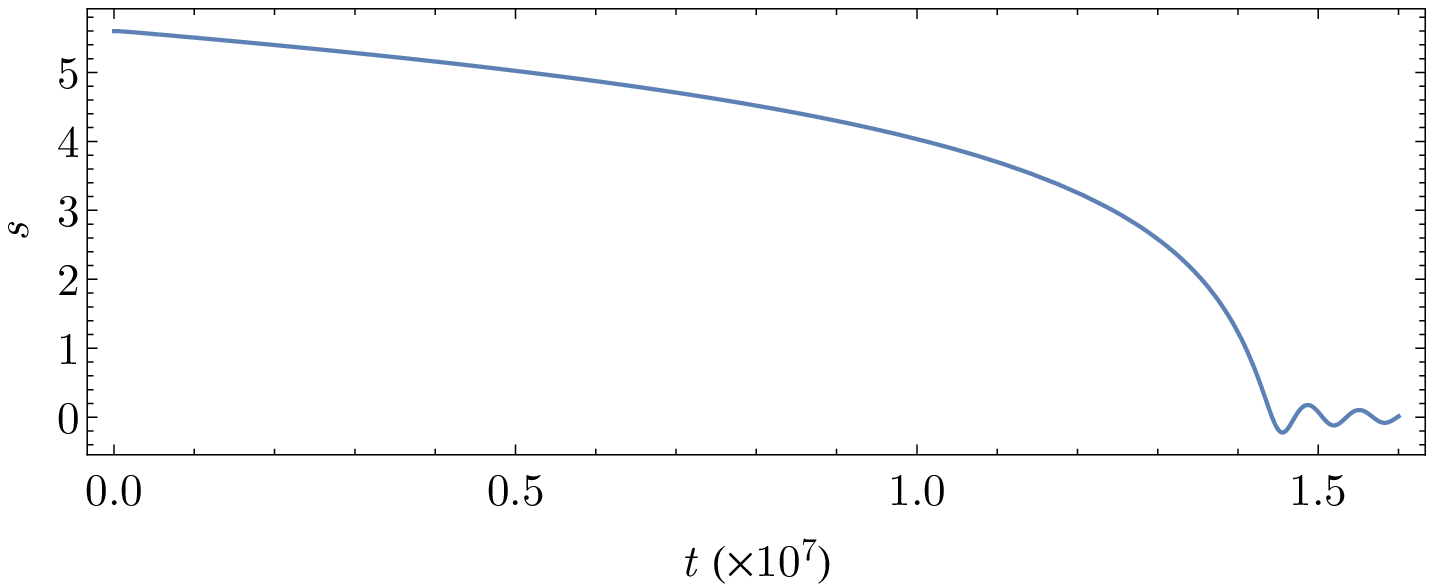}} 
	\vspace{5pt}
	\scalebox{0.70}{\includegraphics{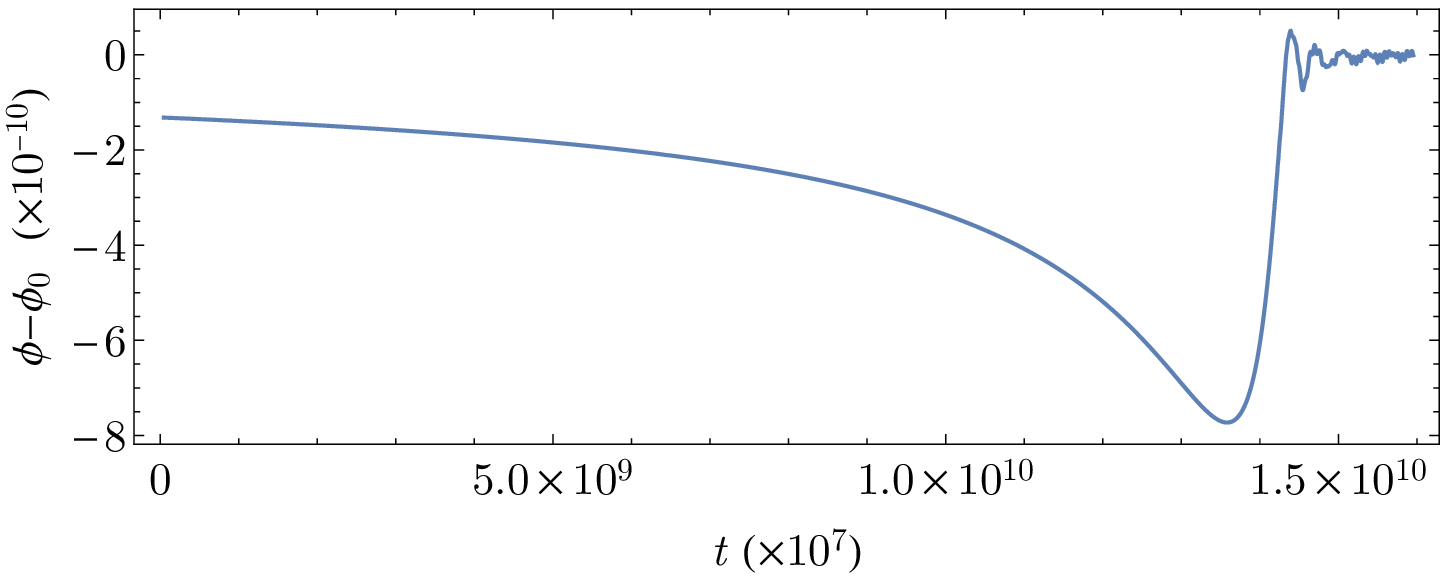}} 
	\vspace{5pt}
	\scalebox{0.70}{\includegraphics{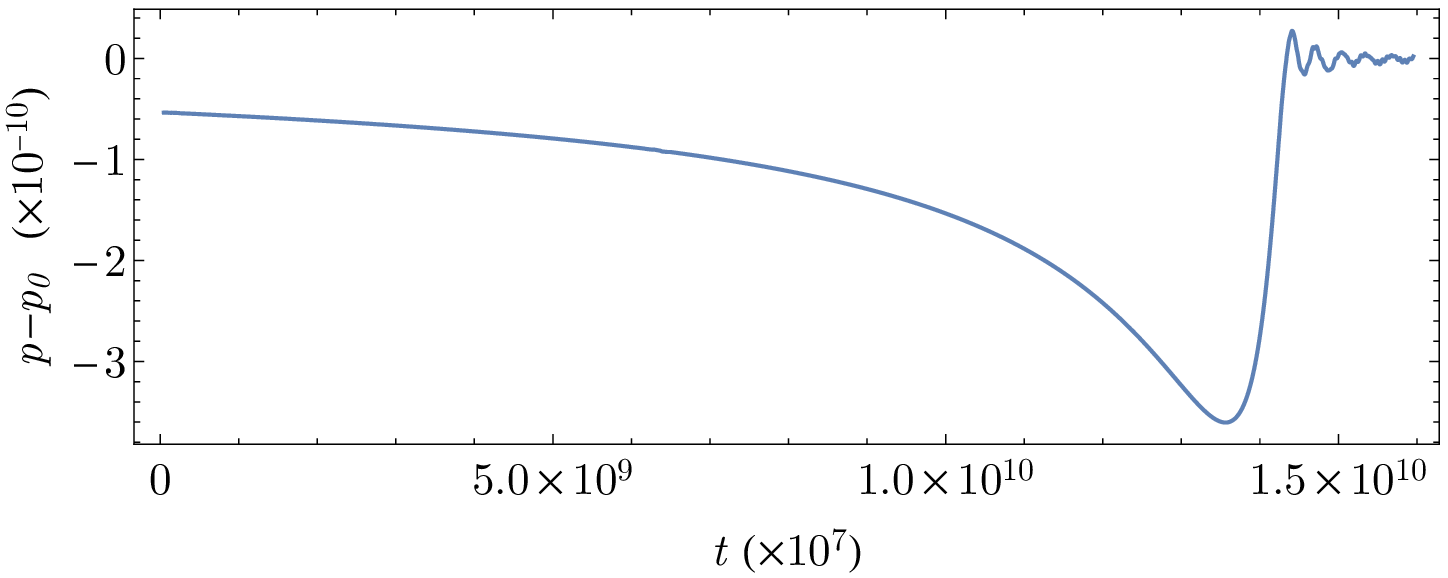}} 
	\vspace{5pt}
	\scalebox{0.70}{\includegraphics{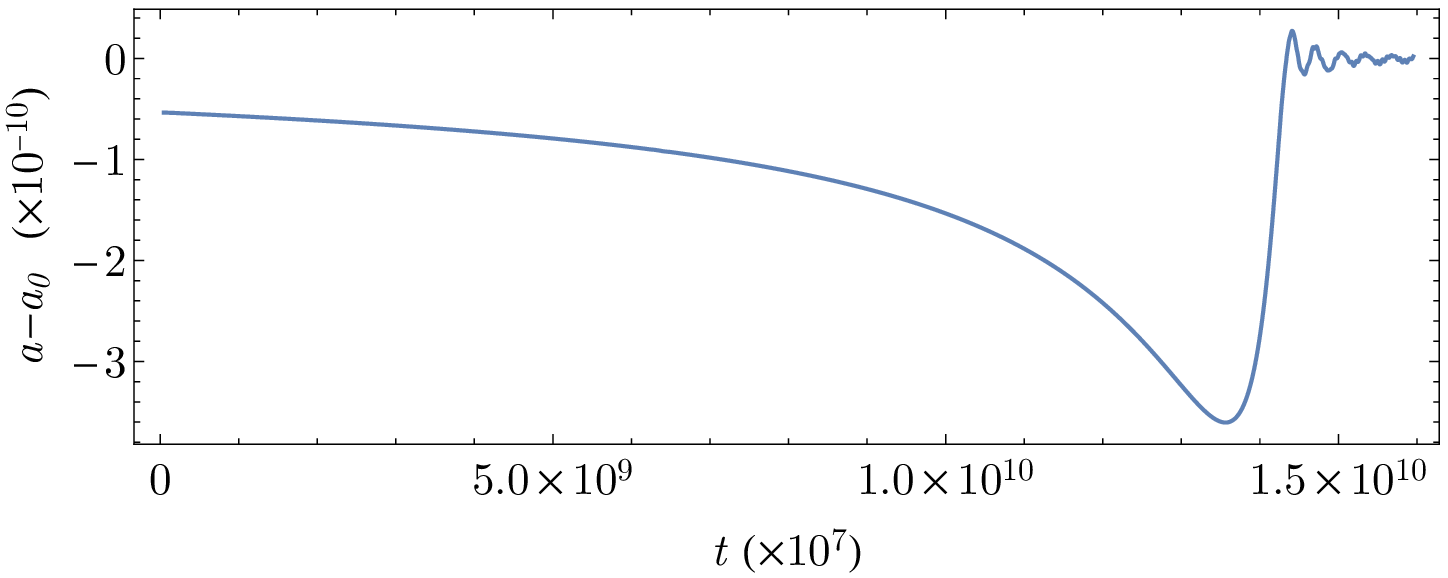}} 
	\scalebox{0.70}{\includegraphics{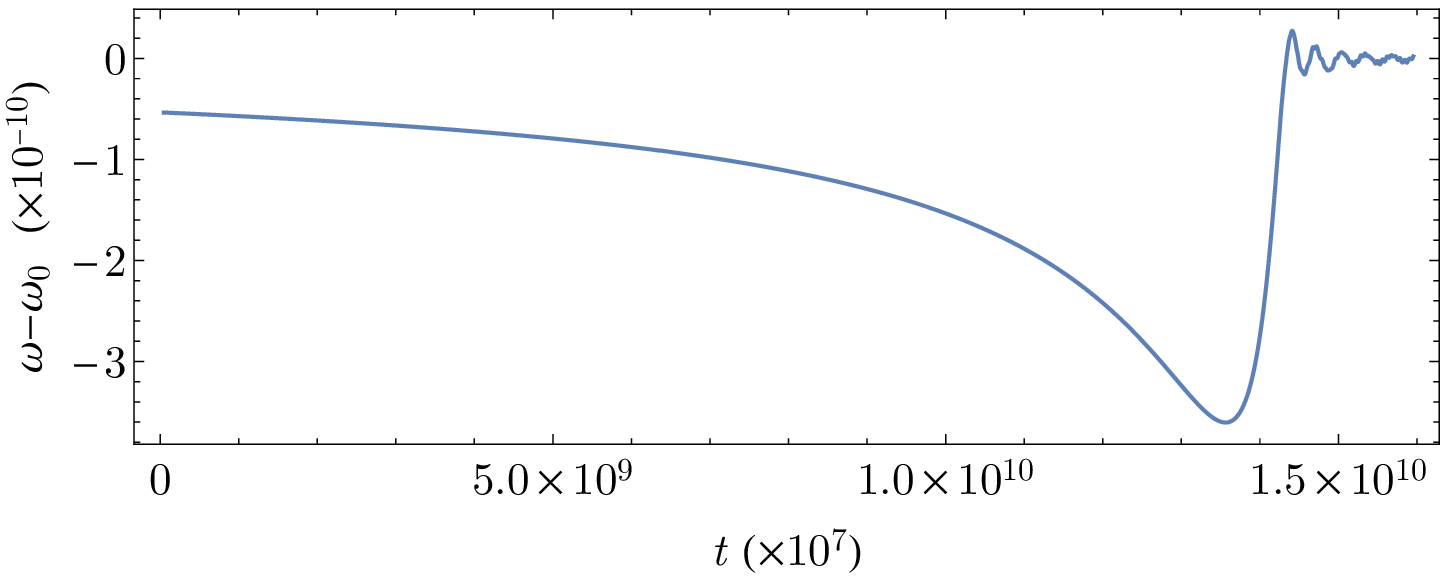}} 
	\caption{\it Evolution of the canonically-normalized inflaton $s$ and the SM singlets $\phi,p,a,\omega$ during inflation, 
	for the parameters (\ref{partpar}) and $x=-1$. The Higgs vevs $\{p_0,a_o,\omega_0,\phi_0\}$ are all equal to $10^{16}\,{\rm GeV}$. For simplicity we display values in Planck units with $M_P=1$.} \label{fig:sol1}
\end{figure}

We consider two types of solutions: 1) $x= -1$ and $\phi = p =a = \omega
= M_{\rm GUT}$; 2) $x \simeq 0$ and $\phi < p, a, \omega$, {\it i.e.},
$M_{\rm int} < M_{\rm GUT}$.  As was discussed previously, for case 1)
the differences between the instantaneous values of the Higgs fields
during inflation and their vacuum vevs are negligibly small, and
inflation can be realized for a wide range of values of $b\ll
1$. Fig.~\ref{fig:sol1} displays the numerical solutions for the SM
singlets $s$, $\phi$, $p$, $a$, $\omega$ during inflation, for the
following set of parameters: 
\beq\label{partpar}
m=10^{-5}\,,\quad m_{\Phi}= 3.3\times 10^{-2}\,, \quad m_{\Sigma}= 8.2\times 10^{-4}\,,\quad  \Lambda=-0.2\,,\quad  \eta=0.8\,,\quad b=10^{-6}\,,
\eeq
with $c =\gamma=0$. These parameter values are chosen to obtain vevs for
the singlet components of the {\bf 210} and {\bf 16} ($\overline{\bf
16}$) equal to $10^{16}\,{\rm GeV}$. The resulting inflationary
parameters are illustrated in Fig.~\ref{fig:planck1} in the range $0\leq
b \leq 3.8 \times 10^{-6}$. As one can see the evolution of all fields
track very smoothly their local minimum as $s$ evolves over the last
$\sim 60$ $e$-folds of inflation. At the end of inflation, all fields
begin oscillations about the low energy vacuum.  

\begin{figure}[!p]
\centering
	\hspace{0.3cm}\scalebox{0.68}{\includegraphics{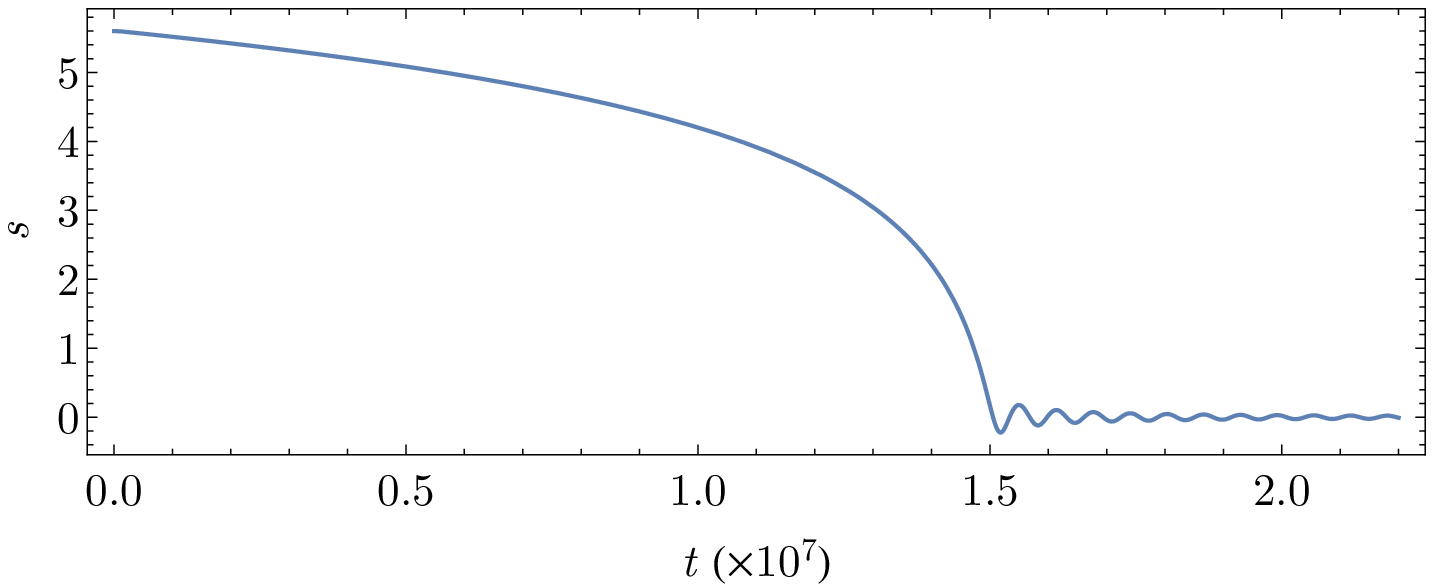}} 
	\vspace{5pt}
	\scalebox{0.70}{\includegraphics{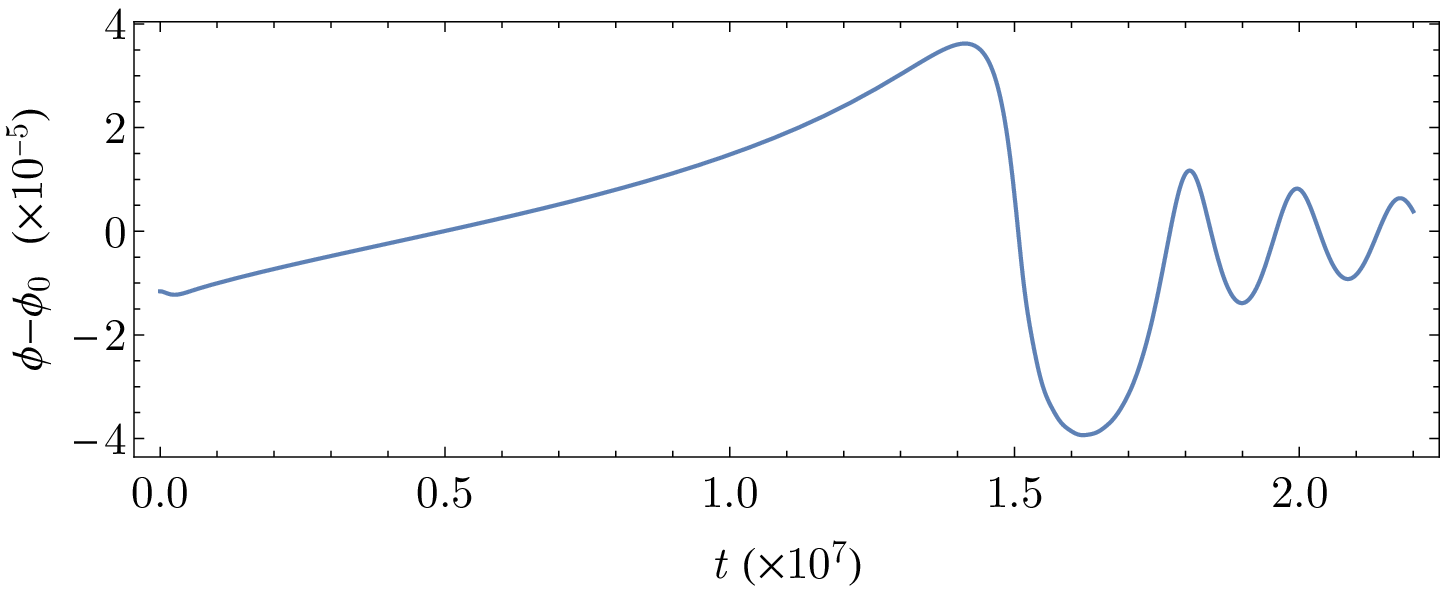}} 
	\vspace{5pt}
	\scalebox{0.70}{\includegraphics{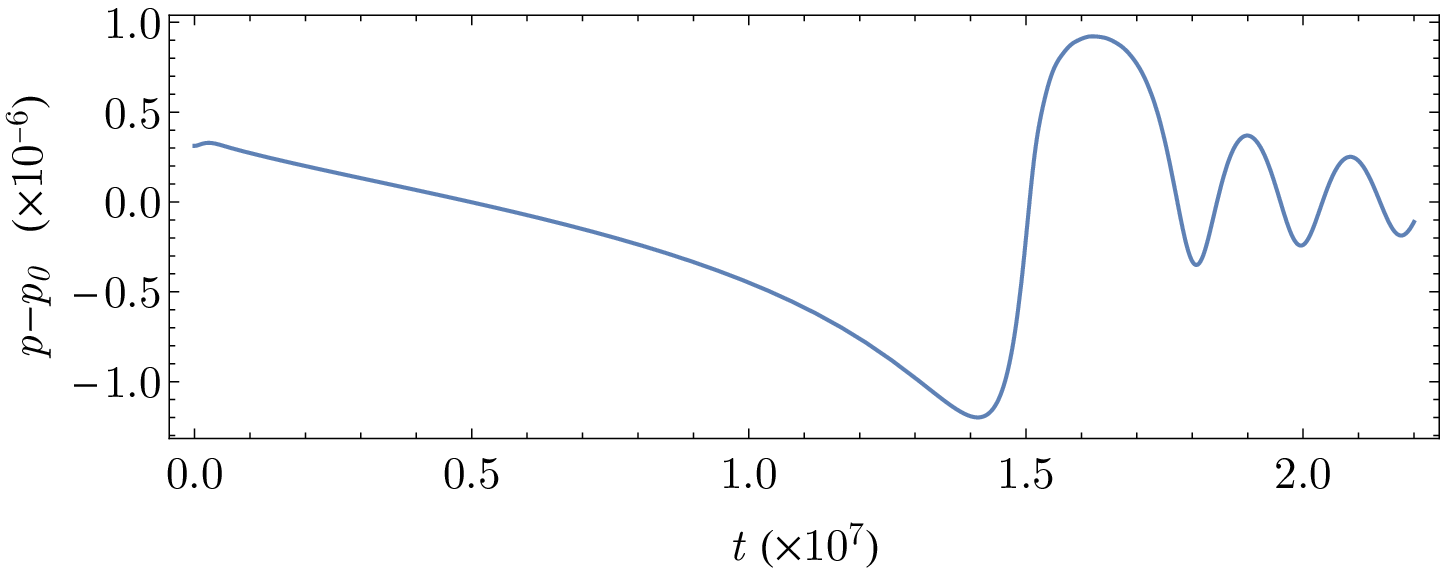}} 
	\vspace{5pt}
	\scalebox{0.70}{\includegraphics{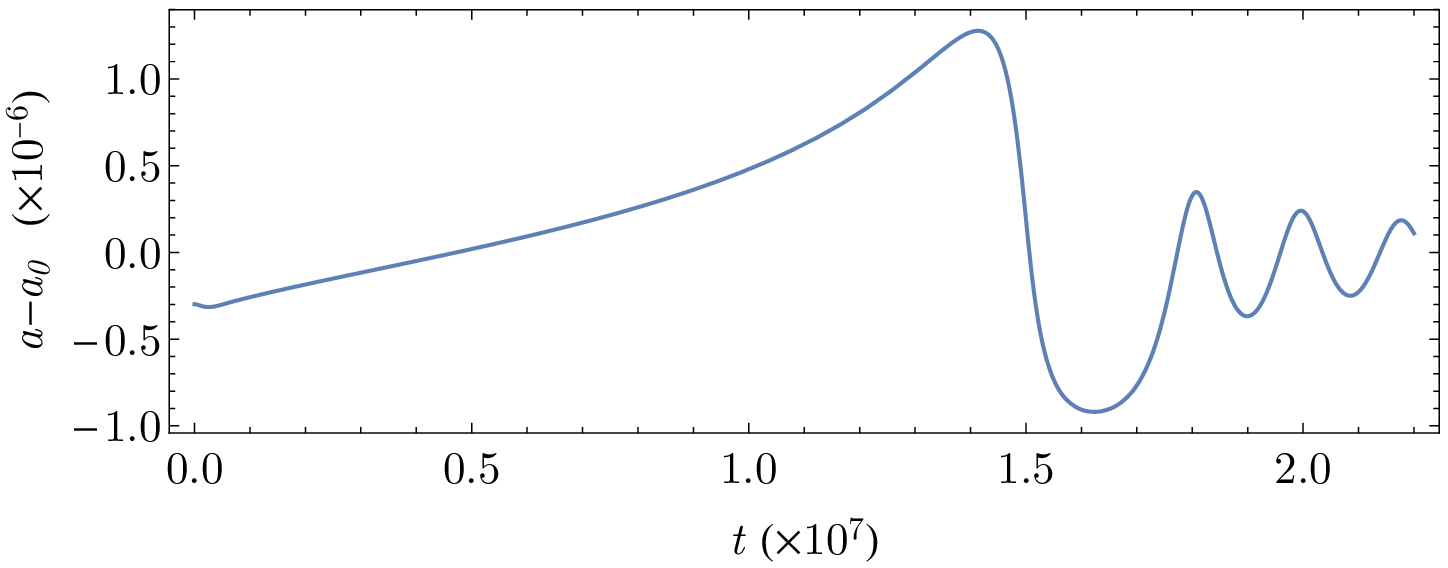}} 
	\scalebox{0.70}{\includegraphics{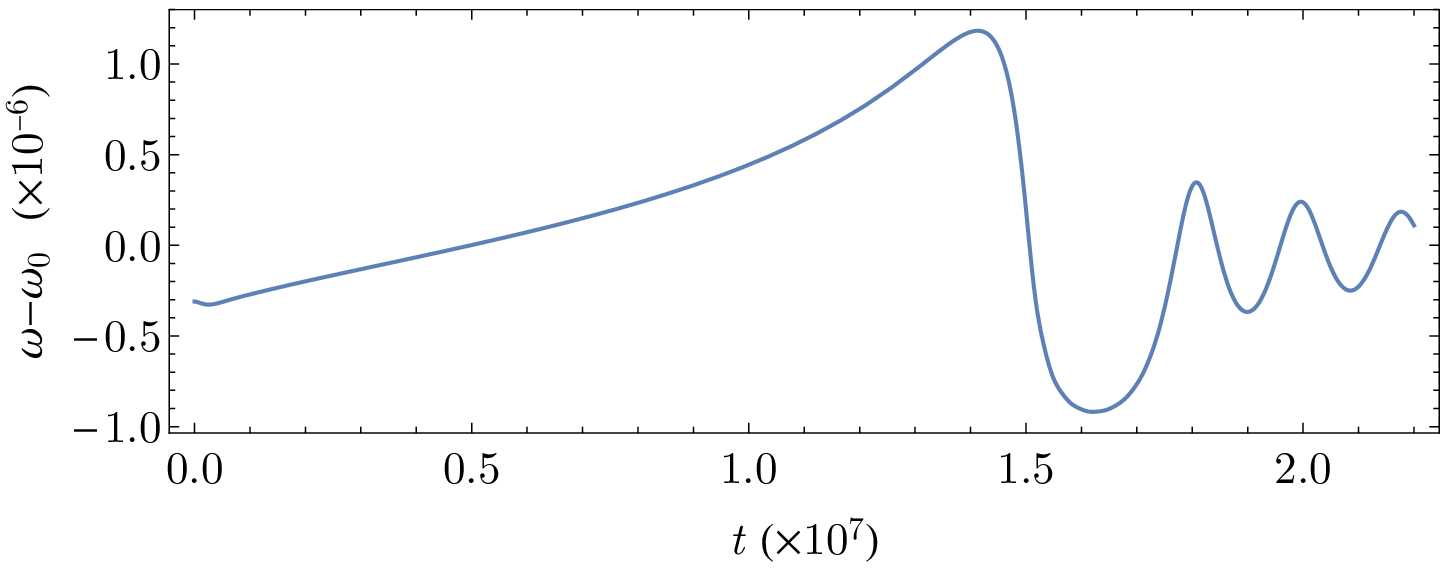}}
	\caption{\it Evolution of the canonically-normalized inflaton $s$ and the SM singlets $\phi,p,a,\omega$ during inflation, 
	for $x=0.0004$. The Higgs vevs are given by $a\simeq -4\times 10^{-3}$, $\phi\simeq 1.1\times 10^{-4}$ and $\omega\simeq -p\simeq 1.6\times 10^{-6}$ in units with $M_P=1$.} \label{fig:sol2}
\end{figure}

As an example of case 2), we consider $x=0.0004$. In this case, as
discussed at the beginning of this Section, we find that the
instantaneous minimum during inflation is displaced relative to its
position at $S=0$ by  $ \delta (p,a,\omega) \propto 10^{-2} b^2/x\eta
m_{\Phi}$ and  $\delta\phi \propto 10^{-2} m_{\Sigma}^{1/2} b^2/ \eta (x
m_{\Phi})^{3/2}$. If $x$ is too small, this deviation can no longer be
considered a perturbation, and it can be shown numerically that the
Higgs fields are driven towards $\phi=0$, $p=a=\omega$, thus eventually
rolling into a ${\rm SU}(5)\otimes {\rm U}(1)$-preserving
minimum~\cite{Bajc:2004xe}. Fig.~\ref{fig:sol2} illustrates a particular
realization of the hierarchy $\langle \Phi\rangle\ll \langle
\Sigma\rangle$ that leads successfully to the SM vacuum. In this case,
the parameters used correspond to $\eta=\Lambda=0$,
$m_{\Sigma}=4\times10^{-4}$, $m_{\Phi}=10^{-3}$ and $b=10^{-6}$. The
vevs in turn correspond to $a\simeq -10^{16}\,{\rm GeV}$, $\phi\simeq
3\times 10^{14}\,{\rm GeV}$ and $\omega\simeq -p\simeq 4\times
10^{12}\,{\rm GeV}$. In this particular case, the Higgs excursions
during inflation are not negligible, which implies that the inflationary
potential does not have the simple form (\ref{Vsimple}), and a numerical
approach must be followed to constrain the value of $|b\phi|$ that would
lead to Planck-compatible inflation. Nevertheless, as
Fig.~\ref{fig:sol2} demonstrates, the bound on $|b\phi|$ does not differ
significantly from the analytical approximation based on
Eq.~(\ref{deltaV}). A smaller value of $x$ would in principle drive the
intermediate scale vev lower, but it can be shown numerically that in
this case the Higgs fields fail to lead to a SM minimum if we choose a
smaller $x$ for any $b\gtrsim 10^{-6}$. We also note in passing that
solutions with small $x$ may be problematic for proton decay as we
discussed above.

\begin{figure}[!p]
\centering
	\hspace{0.3cm}\scalebox{0.68}{\includegraphics{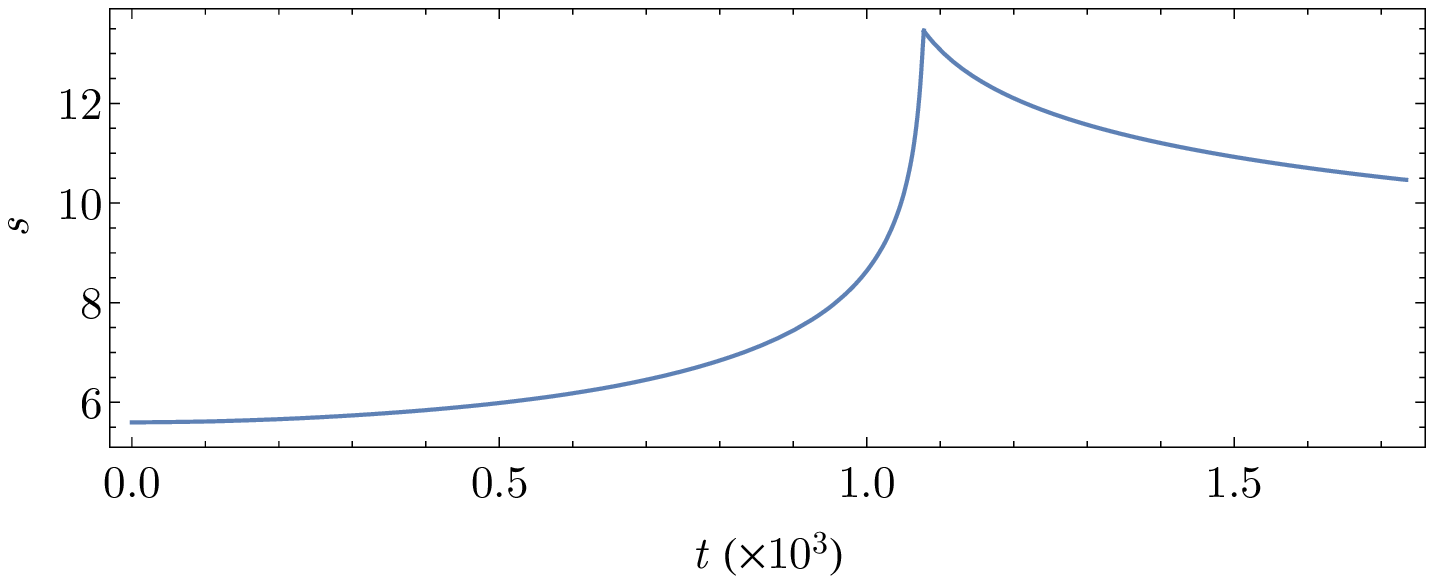}} 
	\vspace{5pt}
	\scalebox{0.70}{\includegraphics{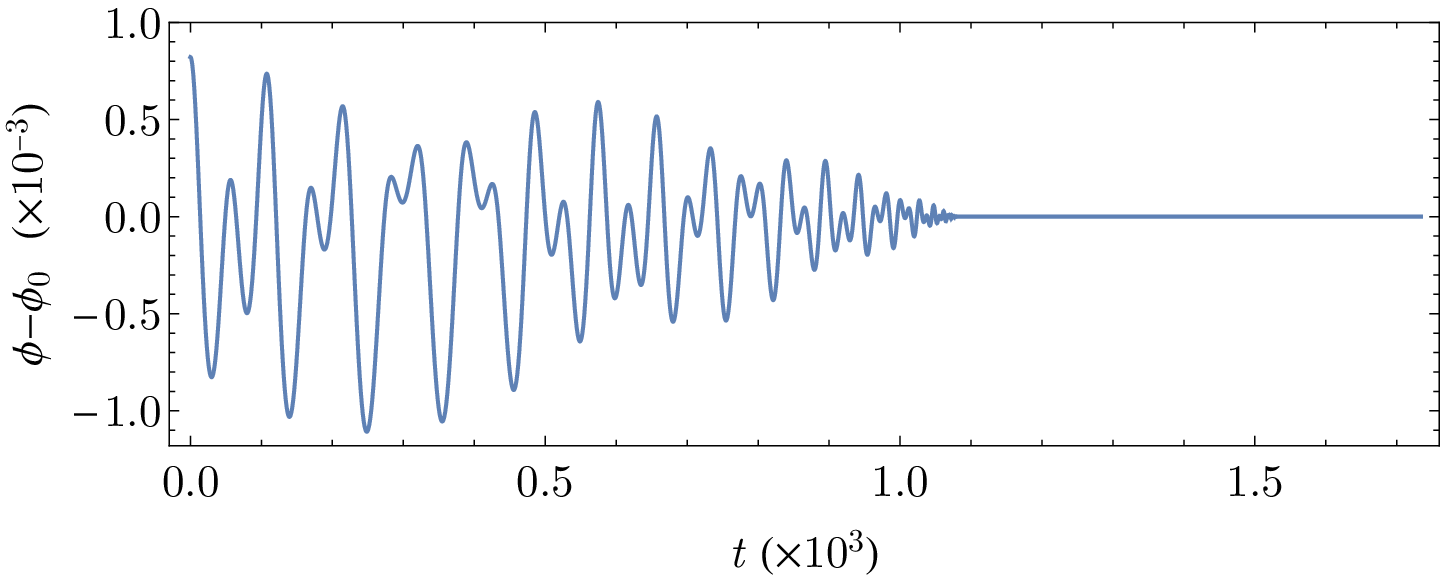}} 
	\vspace{5pt}
	\scalebox{0.70}{\includegraphics{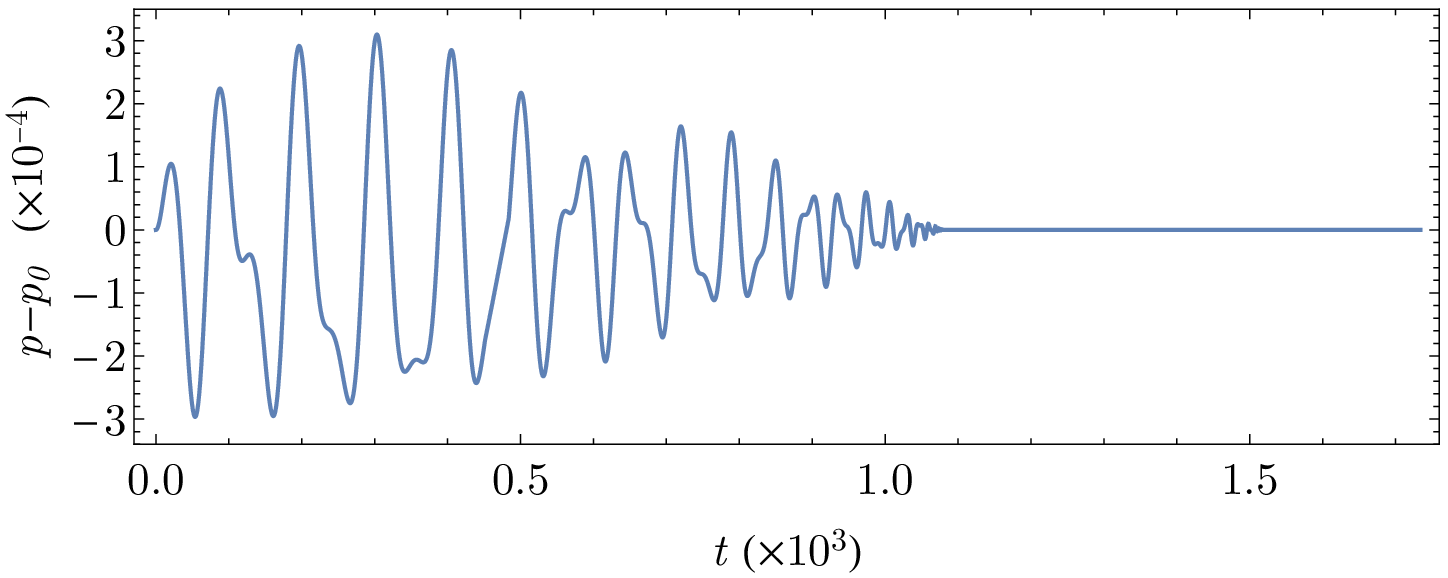}} 
	\vspace{5pt}
	\scalebox{0.70}{\includegraphics{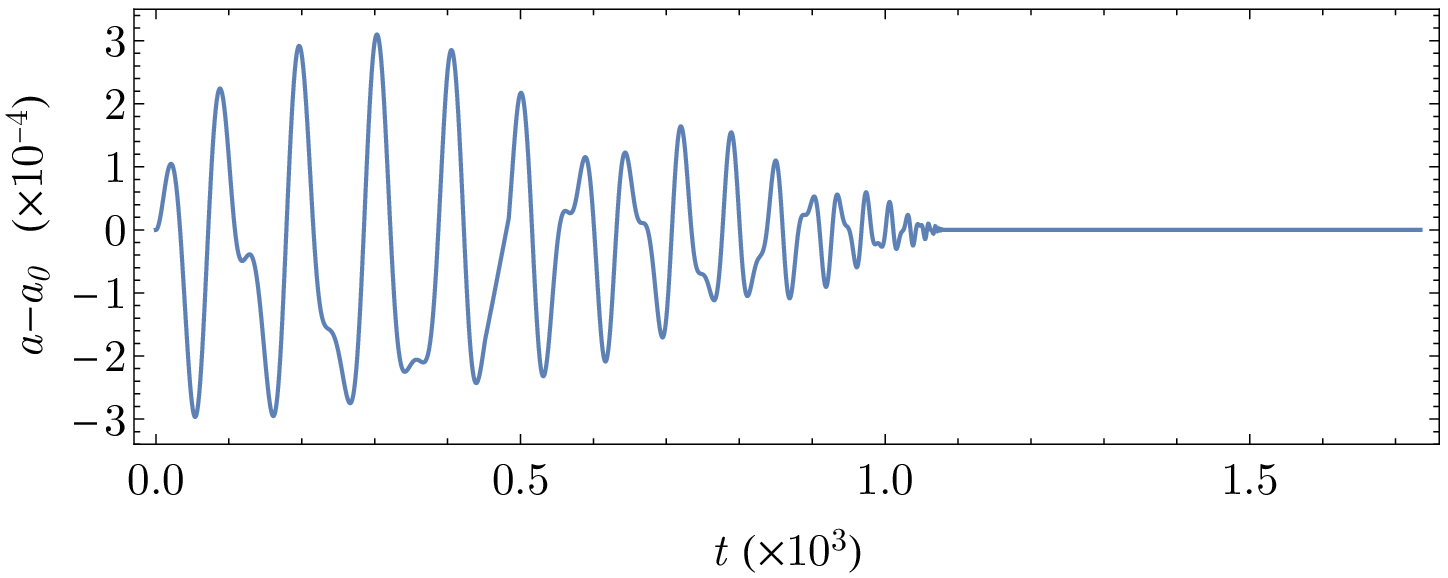}} 
	\scalebox{0.70}{\includegraphics{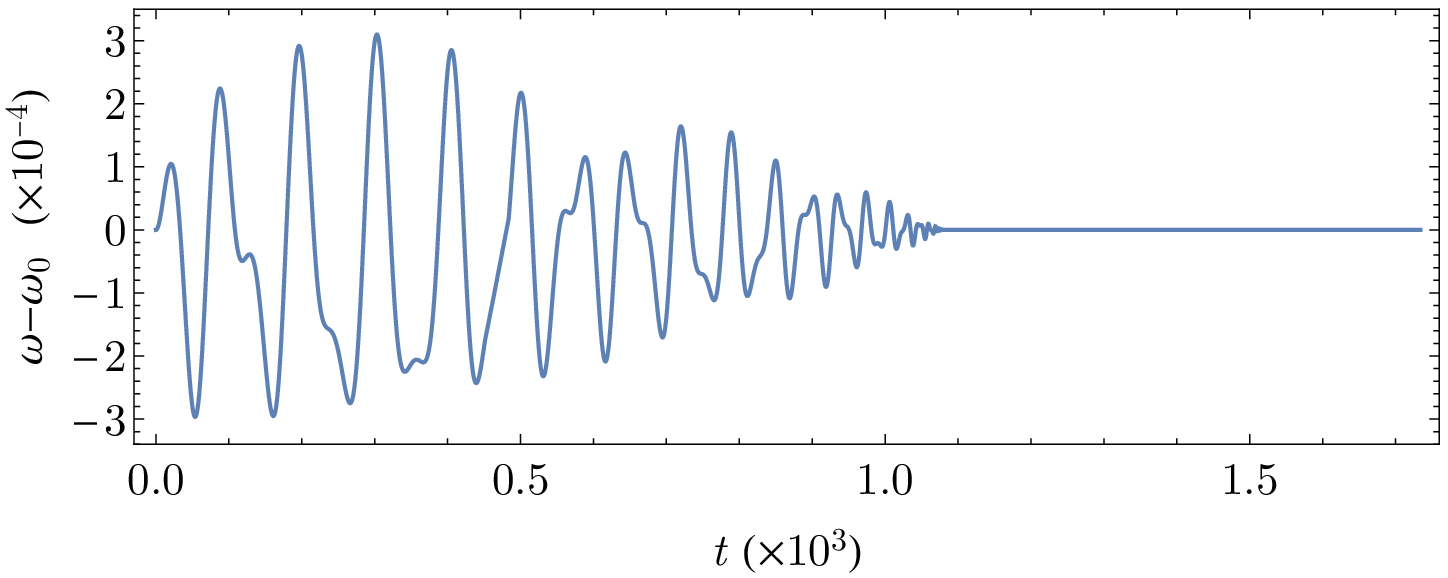}}
	\caption{\it Evolution of the canonically-normalized inflaton $s$ and the SM singlets $\phi,p,a,\omega$ during inflation, 
	for the set of parameters defined by $x=-1$. 
	Here we consider a perturbed initial condition $\Delta\phi/\phi_0=0.2$. Here $\{p_0,a_o,\omega_0,\phi_0\}$ are as in Fig.~\ref{fig:sol2}.} \label{fig:sol1b}
\end{figure}

The initial conditions for the Higgs fields chosen for the numerical
solution shown in Figs.~\ref{fig:sol1}  and \ref{fig:sol2} coincide with
the position of the instantaneous minimum, but we have checked that
inflation and the successive evolution towards the GUT-breaking vacuum
are stable if the initial conditions are perturbed by up to
$\Delta\phi/\phi_0 \lesssim  {\rm few} \times 10^{-1}$. This is
illustrated in Fig.~\ref{fig:sol1b} for an initial deviation
$\Delta\phi/\phi_0 =  0.2$ for case 1 with $x = -1$. We note that the
initial {\em uphill} rolling of the inflaton is seeded by the kinetic
energy of the oscillations of the Higgs fields through the
connection-dependent terms in (\ref{eomfull}), namely: 
\beq
\Gamma^S_{bc}\dot{\Psi}^b\dot{\Psi}^c \simeq
-\frac{1}{2\sqrt{3}}\,\sinh(\sqrt{2/3}\,s) \left(\dot{p}^2+ 2\dot{a}^2 +
6\dot{\omega}^2 + 2\dot{\phi}^2 \right) + \cdots \, . 
\eeq
As the value of $s$ increases, the oscillations of $\phi$ are rapidly
damped, and the subsequent evolution resembles that shown in
Fig.~\ref{fig:sol1}. Note the difference in timescale in this
figure. The transient growth in $s$ implies an increased total number of
$e$-folds compared to an unperturbed initial condition. Similarly, for
case 2 we are not required to fine-tune the initial positions of the
fields with respect to their minima. However, if these perturbations are
initially too large, the subsequent evolution may well take the theory
to an SO(10)-symmetric vacuum.

So far we have neglected the effects of the couplings $c$ and $\gamma$,
which are not independent as they satisfy the relation
(\ref{eq:vevsing}). Let us for simplicity assume that $c\ll 1$ and
$b=0$. In this case, for $s\gg 1$, the Higgs singlet components are
displaced from their vevs during inflation by corrections that depend
linearly on $c$; for example, 
\beq
\delta \phi \simeq \frac{c}{\eta}\left(\frac{m_{\Sigma}}{m_{\Phi}}\right)^{1/2}\,g(x)\,,
\eeq
where 
$g(x)$ is another (somewhat complicated and long) function of
$x$. Similarly to the $b\neq 0$, $c=0$ case, the function $g(x)$ is
divergent for $x=0,1/3$ and $\pm i$, implying that $\phi$ will be always
driven to zero for $x$ close to these points. For any $x$, with a
sufficiently large $c$, the corrections will be large due to the induced
mass-squared $\sim (cS)^2$, and {\em all} Higgs singlets will be driven
to zero during inflation, leaving the universe in an SO(10)-symmetric
state. In the particular case with $x=-1$ and
$\phi=p=a=\omega=10^{16}\,{\rm GeV}$, this occurs for $c\gtrsim 3\times
10^{-3}$. 

As the analytic approximation (\ref{Vbcg}) is valid for very small $c$,
one would be tempted to relate directly the Planck constraint on
$|b\phi|$ with a constraint on the combination $2|c\phi|^2+|2\gamma p|^2
+ |6\gamma a|^2 + |12\gamma\omega|^2$. However, it can be verified
numerically that values of $c$ larger than the value that one would
naively have expected to be the maximum compatible with Planck data can
still lead to Planck-compatible results; the deviations of the fields
with respect to their vevs compensate the expected deformation of the
inflaton potential. For example, in the previously-discussed $x=-1$
case, the naive expectation would result in the bound $c\lesssim 5\times
10^{-7}$, whereas a numerical calculation shows that 95\% Planck
compatibility is retained for $c\lesssim 7\times 10^{-4}$, only a factor
of four below the maximum value of $c$ allowed by symmetry breaking. 

The specific limits on $b$ and $c$ when both are non-vanishing must be
checked numerically on a case-by-case basis. Nevertheless, it is clear
that the allowed values of $b$ and $c$ are reduced due to the
simultaneous effect of both couplings.

\section{Yukawa Couplings and Neutrino Masses}
\label{sec:neutrinomass}

\subsection{Yukawa unification and its violation}

As discussed in Section~\ref{sec:dtspl}, the MSSM Higgs fields $H_u$ and
$H_d$ in our model are given by linear combinations of the SU(2)$_L$
doublet components in the fields $\Phi$, $\bar{\Phi}$, and $H$. The
Yukawa coupling terms in the low-energy effective theory are then written as
\begin{equation}
 W_{\rm Yukawa} = f_u H_u Q \bar{u}
+ f_\nu H_u L {\nu}^c_R
- f_d H_d Q\bar{d}
- f_e H_d L\bar{e} ~,
\label{eq:wyukawa}
\end{equation}
where the Yukawa couplings are related to the corresponding GUT Yukawa
couplings through the following GUT-scale matching conditions: 
\begin{equation}
 f_u = f_\nu = y\, {\cal U}_{11} ~,~~~~~~
 f_d = f_e =y{\cal D}_{11} ~.
\label{eq:gutcond}
\end{equation}
These equations show that we expect the unification of down-type quark
and charged-lepton Yukawa couplings at the GUT scale, as in SU(5)
GUTs, and the up-type quark and neutrino
Yukawa couplings are also unified. These two classes of the Yukawa
couplings may, however, be different from each other if $\alpha \neq
\bar{\alpha}$, since ${\cal U}_{11} \neq {\cal D}_{11}$ in this
case. This feature distinguishes our model from other SO(10) GUT models,
where one usually has $f_u = f_d = f_e = f_\nu$ at the GUT scale.

These GUT relations are modified if there exist higher-dimensional
operators suppressed by the Planck scale \cite{nro}. Among such
operators, the following dimension-five operator is expected to give the
leading contribution: 
\begin{equation}
 W_{\rm eff} = \frac{c_{\Delta f}}{M_P} H \Sigma \psi \psi ~.
\label{eq:weff5}
\end{equation}
After $\Sigma$ develops a vev, this operator leads to the Yukawa
couplings in Eq.~\eqref{eq:wyukawa}. The matching conditions in this
case are given by
\begin{align}
 f_u &= (y + \Delta f)\,{\cal U}_{11} ~, ~~~~~~
 f_\nu = (y -3 \Delta f) \,{\cal U}_{11} ~,
\nonumber \\
 f_d &= (y + \Delta f)\,{\cal D}_{11} ~, ~~~~~~
 f_e = (y -3 \Delta f) \,{\cal D}_{11} ~,
\end{align}
with
\begin{equation}
 \Delta f = \frac{c_{\Delta f}}{M_P} (a + \omega) ~.
\end{equation}
Since $\Delta f = {\cal O}(10^{-3})$ for $ c_{\Delta f} = {\cal O}(1)$, 
the GUT relations for the first- and second-generation
Yukawa couplings may be modified significantly in the presence of the
dimension-five operator. For the third-generation Yukawa couplings, on
the other hand, its effects are less significant. Intriguingly, this is
consistent with the observed quark and lepton mass spectrum; experimentally, bottom
and tau Yukawa unification is realized at the ${\cal O}(10)$\%
level in most of the parameter space in the MSSM~\footnote{The corresponding
relation in non-supersymmetric SU(5) GUT actually led to a successful prediction of the $b$ quark
mass before its discovery: see the third paper in~\cite{so10}.}, while the deviations
in $s$-$\mu$ and $d$-$e$ unification are as large as ${\cal
O}(100)$\%.

\subsection{Neutrino masses}

We now investigate the mass matrix for neutrinos. If we take $M, b^\prime,
\alpha^\prime \to 0$, then $\Phi$ and $\bar{\Phi}$ have no mixing with
neutrinos; we consider this limit for simplicity. Note that this limit
suppresses the $R$-parity violating operators, and thus is
phenomenologically desirable as we discuss below. A non-zero value of the
coupling $b$ induces mixing between right-handed neutrinos and the
singlinos $\tilde{S}_i$, which are the fermionic component of the
singlet superfields $S_i$. We also suppress the couplings $c$, $\gamma$,
and $\lambda_{SH}$ in order to prevent $S$ from mixing with $\Phi$,
$\bar{\Phi}$, $\Sigma$,  and $H$. As we have seen above, for the
inflaton field, the smallness of $c$ and $\gamma$ is required by
successful inflation, while $\lambda_{SH}$ should be small in order to
avoid over-production of gravitinos as we will see in the next
section. In this case, mixing occurs only among the right- and
left-handed neutrinos and the singlinos. Disregarding Planck-suppressed
factors, the neutrino-singlino fermion mass matrix takes the form
\cite{GN}  
\beq \label{seeM}
{\cal L}_{\rm mass} = -
\begin{pmatrix}
\overline{\nu}_L & \overline{\nu}^c_R & \overline{\tilde{S}} 
\end{pmatrix}
\begin{pmatrix}
0 & -f_\nu\,v\sin\beta & 0 \\
-f_\nu\,v\sin\beta & 0 & -b\phi\\
0 & -b\phi & m
\end{pmatrix}
\begin{pmatrix}
\nu_L \\[3pt]
\nu_R^c\\[3pt]
\tilde{S}
\end{pmatrix}
 \, ,
\eeq
where $v\simeq 174$~GeV is the Standard Model Higgs vev and $\tan \beta
\equiv \langle H_u\rangle /\langle H_d \rangle$. A similar form for the
mass matrix is found in flipped SU(5) \cite{flipped2,ENO3}.

For the first-generation neutrinos, the requirement of successful
inflation restricts the coupling $b$ as we have seen in the previous
section. In this case, the couplings satisfy the hierarchy
\beq
f_\nu\,v\sin\beta \ll b\phi \ll m ~,
\eeq
and thus the diagonal mass matrix has a double-seesaw form given
approximately by 
\beq\label{numasses}
{\cal L}_{\rm mass} = -
\begin{pmatrix}
\overline{\nu}_L^M & \overline{\nu}_R^M & \overline{\tilde{S}}^M 
\end{pmatrix}
\begin{pmatrix}
m\left(\frac{f_\nu\,v\sin\beta}{b\phi}\right)^2 & 0 & 0 \\
0 & -\frac{(b\phi)^2}{m} - m\left(\frac{f_\nu\,v\sin\beta}{b\phi}\right)^2 & 0\\
0 & 0 & m + \frac{(b\phi)^2}{m}
\end{pmatrix}
\begin{pmatrix}
\nu_L^M \\[5pt]
\nu_R^M\\[5pt]
\tilde{S}^M
\end{pmatrix}
 \, ,
\eeq
and the corresponding mass eigenstates are approximately
\begin{align} \label{nuM1}
\nu_L^M \;&\simeq\;  \nu_L \,-\, \frac{m\,f_\nu v\sin\beta}{(b
 \phi)^2}\,\nu_R^c \,-\, \frac{f_\nu v\sin\beta}{b \phi}\,\tilde{S} \, , \\
\nu_R^M \;&\simeq\;  \nu_R^c \,+\, \frac{b\phi}{m}\,\tilde{S} \,+\,
 \frac{m\,f_\nu v\sin\beta}{(b\phi)^2}\,\nu_L \, , \\ \label{nuM3}
\tilde{S}^M \;&\simeq\;  \tilde{S} \,-\, \frac{b\phi}{m}\,\nu_R^c \,+\,
 \frac{b \phi \,f_\nu v\sin\beta}{m^2}\,\nu_L \, .
\end{align}
For the second and third generations, on the other hand, the coupling
$b$ (recall we have suppressed all generation indices) 
can be arbitrary, but the masses for light neutrinos are still given
by
\begin{equation}
 m_\nu \simeq m\left(\frac{f_\nu\,v\sin\beta}{b\phi}\right)^2 ~.
\end{equation}

\begin{figure}[ht]
\centering
    \scalebox{0.85}{\includegraphics{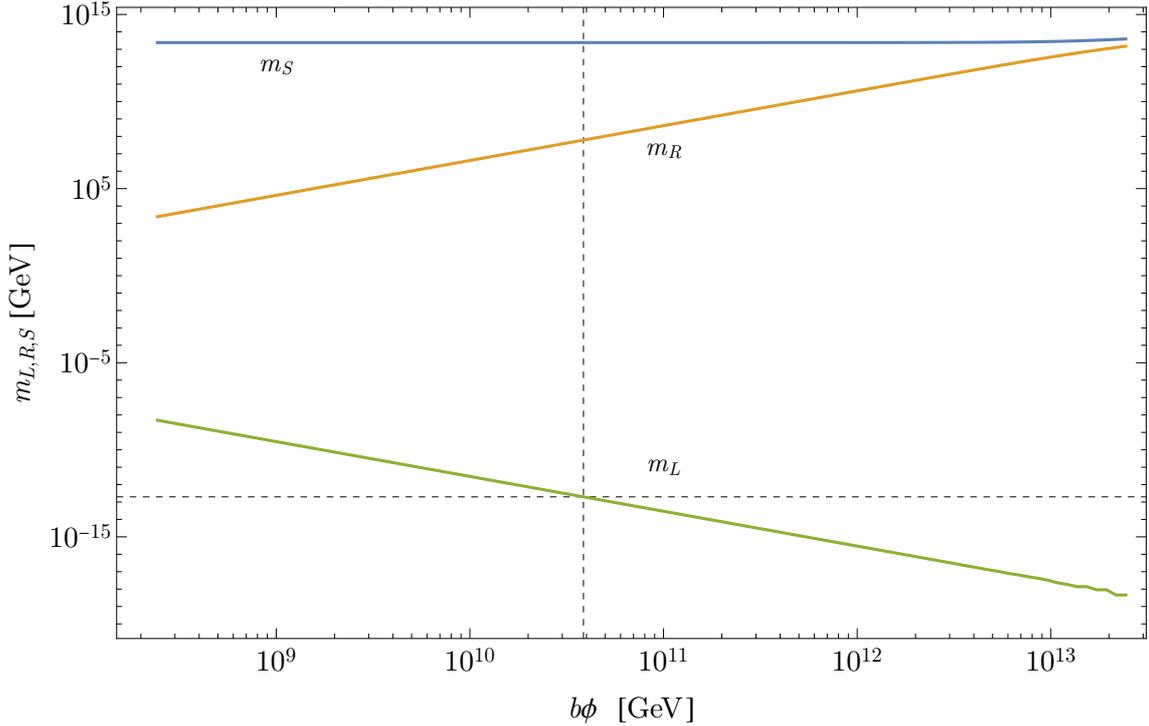}}
    \caption{\em The neutrino mass spectrum in the double-seesaw
 scenario arising from (\ref{seeM}), assuming $m=10^{-5}M_P$ and $f_\nu
 \sin\beta=2\times 10^{-5}$, as a function of $b \phi$. For these
 parameters, the allowed range for $b\phi$ and correspondingly the
 neutrino masses, is located in the upper left region, bounded to the
 left by the vertical dashed line showing the 95\% CL Planck upper limit
 $b\phi=10^{-7.8}M_P$, and below by the horizontal dashed line
 corresponding to $m_{L}\simeq 2\times 10^{-13}\,{\rm GeV}.$ } 
    \label{fig:masses1}
\end{figure}

Values of the mass eigenvalues calculated numerically for $m\simeq 10^{-5}M_P$
are shown in Fig.~\ref{fig:masses1}. If $b\phi \sim 10^{-8.5}M_P$, a
left-handed neutrino mass $m_{L}\lesssim 0.1\,{\rm eV}$ is obtained if
$f_\nu \sin\beta \lesssim 9\times 10^{-5}$~\footnote{We note that the up-quark
mass $m_u = 2.3$~MeV \cite{Agashe:2014kda} and the GUT relation
\eqref{eq:gutcond} implies $f_\nu \simeq 10^{-5}$, which is consistent
with the above limit.}. Currently, the Planck 2015 data
\cite{planck15} imposes a constraint on the sum of the
neutrino masses: $\sum m_{\nu} < 0.23$~eV, and an even stronger
bound of $\sum m_{\nu} < 0.12$~eV 
comes from the Lyman $\alpha$ forest power spectrum obtained by BOSS
in combination with CMB data~\cite{15A}.   
Since the parameters $b$ and $m$ for the second
and third generations are almost arbitrary, we may explain the current
neutrino oscillation data \cite{Capozzi:2016rtj} by appropriately
choosing these parameters (as well as flavor-changing couplings
corresponding to $b$ and $m$). For $f_\nu \sin \beta = m_u/v$ and
$m= 10^{-5}$, the normal hierarchical neutrino mass spectrum is
obtained if $b\phi \gg 10^{-9}$. On the other hand, the neutrino mass spectrum is inverted
if $b \phi \simeq 7 \times 10^{-10}$. A
smaller value of $b\phi$ leads to a quasi-degenerate mass spectrum. 
The latter two types of mass spectrum are constrained
by both neutrino oscillation data \cite{Capozzi:2016rtj} and the CMB
observations \cite{planck15, 15A}, and will be tested in future
experiments. It has not escaped our attention that there is a strong correlation
between the CMB observations, as quantified in the values of $n_s$ and $r$, and the light neutrino masses,
that becomes apparent if we write (\ref{nsan}) and (\ref{ran}) as
\begin{align}
n_s &\;\simeq\; -\frac{2}{N_*} + \frac{8}{3}\left(\frac{m}{m_{\nu}}\right)\left(\frac{f_{\nu}\,v\sin\beta}{m}\right)^2N_*^2 + \frac{32}{81}\,\Delta K\,N_*\,,\\
r &\;\simeq\; \frac{12}{N_*^2}  + \frac{32}{3}\left(\frac{m}{m_{\nu}}\right)\left(\frac{f_{\nu}\,v\sin\beta}{m}\right)^2 N_* + \frac{64}{27}\,\Delta K\,.
\end{align}

When the couplings $M$ and $b^\prime$ are different from zero (and
related by the minimization condition (\ref{eq:vevnur})), the fermion
mass matrix for the SM singlets and uncharged doublets ceases to be
block-diagonal, and potentially large terms such as
$\mathcal{M}^{\nu_L\bar{\phi}_L}=M+b^\prime(p-3a)$ or
$\mathcal{M}^{\nu_R\,p}=b^\prime\phi$ will in general result in
significantly mixed mass eigenstates. As a crude approximation, if one
assumes that only the `left-handed' fields mix, {\it i.e.}, the fermionic 
components of $H_L,\bar{H}_L,\phi_L,\bar{\phi}_L$ together with $\nu_L$,
one can compute, {\it e.g.}, the contribution of the $\tilde{\phi}_L$
gauge eigenstate to the lightest state, which in the
$b^\prime\rightarrow 0$ limit would correspond to a pure $\nu_L$
state. This contribution has the form 
\beq
\psi_{\rm lightest}^{(\phi_L)} \simeq \frac{m_{H}}{\Delta}(M+b^\prime(p-3a))\,,
\eeq
where $\Delta\ll M_{\rm GUT}^2$ has been defined in (\ref{eq:deltadef}),
and is related to the weak-scale $\mu$-term via $\Delta\simeq
\mu\,[m_H+2\eta(p+3\omega)]$. This implies that sizable mixing can occur
for $b^\prime \gtrsim \mu/M_{\rm GUT}$. For a larger $b^\prime$ (and
thus $M$), the fine-tuning condition for the doublet-triplet splitting
is modified, and it turns out that $\Delta$ should also grow ($\Delta
\propto b^\prime$ for $b^\prime \gg \mu/M_{\rm GUT}$)
to keep $\mu$ at the supersymmetry-breaking scale. In any case, as we
see in the subsequent section, we need to $b^\prime, M$ to be small in
order to ensure a good dark matter candidate (the lightest neutralino
with a lifetime longer than the age of the Universe), and thus we do not
consider further in this paper the case of large $b^\prime, M$.

As discussed previously, non-vanishing values of $c$ and $\gamma$ would mix
the Higgs sector and the inflatino $\tilde{S}$. However, we have
verified that, in the Planck-allowed range for $c$, the mass spectrum,
and in particular the left-handed neutrino state, are negligibly
affected. 

\section{Reheating and Leptogenesis}
\label{sec:reheating}

In the absence of a direct coupling between the inflaton and matter,
reheating in supergravity models almost always proceeds through the minimal
gravitational couplings \cite{nos}, leading to a minimal reheat temperature
of order $10^6$~GeV \cite{Endo:2006qk}. However, these couplings
vanish in no-scale supergravity~\cite{ekoty} and reheating must proceed
either though a direct coupling to matter or a coupling to gauge fields
through the gauge kinetic term. For this reason, the identification of
the inflaton with the right-handed sneutrino has appeared to be very
promising, as reheating takes place naturally through the decays of the
inflaton to sneutrino/Higgs or neutrino/Higgsino pairs
\cite{ENO8,EGNO4}. In fact, to avoid excessive reheating and gravitino
production, it was necessary to set a limit on Yukawa coupling of the
inflaton (right-handed sneutrino) of order $10^{-5}$, comparable to the
electron Yukawa coupling. 

In the present context, the inflaton is once again a singlet, and the
coupling $b$ yields the direct coupling of the inflaton to Standard
Model matter fields through the $\bar{\Phi}$-$H$ and neutrino-singlino
mixings. For the former, the $bS\bar{\Phi}\psi$ term leads to $b\, {\cal
U}_{21} SH_u L$ via the mixing \eqref{eq:udmixing}. For the latter, 
the neutrino Yukawa coupling $f_\nu$ induces an inflaton-Higgs-neutrino
coupling through the scalar mixing
\begin{align}
\tilde{\nu}_R &\simeq \tilde{\nu}_R^{M} - \frac{b\phi}{m}\, S^{M}\, , \\
S &\simeq S^{M} + \frac{b\phi}{m}\,\tilde{\nu}_R^{M}\,,
\end{align}
where we have disregarded weak-scale terms, cf.,
Eqs.~(\ref{nuM1}--\ref{nuM3}). As a result, we obtain an interaction
\begin{equation}
 {\cal L}_{\rm int} = - C_{SHL}\, S^M H_u L ~,
\label{scouple}
\end{equation}
with
\begin{equation}
C_{SHL} =  b\left({\cal U}_{21} -\frac{f_\nu\phi}{m}\right) ~.
\end{equation}
This results in the inflaton decay rate
\beq
\Gamma(S\rightarrow H_u\tilde{L}) + \Gamma(S\rightarrow \tilde{H}_u L) =
\frac{m}{4\pi}\left|C_{SHL}\right|^2\,,
\eeq
which leads to a reheat temperature 
\beq
T_R \simeq 10^{15} ~{\rm GeV}\times
\left|C_{SHL}\right|\left(\frac{g_*}{915/4}\right)^{-1/4}\left(\frac{m}{10^{-5}M_P}\right)^{1/2}
~,
\eeq
where $g_*$ denotes the effective number of degrees of freedom, and $g_*=915/4$ for the MSSM.

The abundance of gravitinos is determined by the 
reheat temperature~\cite{thermal}:
\beq
\frac{n_{3/2}}{s} \simeq 2.4 \times 10^{-12} \left( \frac{T_R}{10^{10} ~{\rm GeV}} \right) \, ,
\eeq
where $s$ is the entropy density and we have assumed that the gravitino is much heavier than the gluino. 
In order to satisfy the upper limit on the abundance of neutralinos: $\Omega_\chi h^2 < 0.12$, we must ensure that~\cite{ego}
\beq 
\frac{n_{3/2}}{s} \lesssim 4.4 \times 10^{-13} \left( \frac{1 {\rm TeV}}{m_\chi} \right) \, ,
\eeq
which leads to an upper limit on the coupling $C_{SHL}$:
\beq
\left|C_{SHL}\right|\lesssim 10^{-5}\,.
\eeq
Since we expect ${\cal U}_{21} ={\cal O}(1)$, $f_\nu ={\cal
O}(10^{-5})$, and $m/\phi ={\cal O}(10^{-3})$ in our setup, this bound
implies $|C_{SHL}| \sim |b| \lesssim 10^{-5}$. 
For a more detailed discussion of reheating, see~\cite{Ellis:2015jpg}. 



If $\lambda_{SH} \neq 0$, the inflaton can decay into a pair of
higgsinos as well. Similarly to the above case, to evade over-production of
gravitinos, we need to suppress this coupling such that $|\lambda_{SH}|
\lesssim 10^{-5}$.

As noted earlier, $R$-parity is violated in this model though, as
described in \cite{ENO8}, the violation via the coupling $b$ is weak
enough to ensure that the lifetime of the lightest supersymmetric
particle is much longer than the age of the Universe. To stabilize the
lightest supersymmetric particle, we need to take $\alpha^\prime$, $M$
and $b^\prime$ to be zero since they make it decay at tree level. 
The form of the coupling of $S$ in Eq.~(\ref{scouple})
is clearly an $L$-violating decay, so the reheating process
may well lead to a lepton asymmetry given by
\cite{dl,nos} 
\beq
\frac{n_L}{s}  \sim \epsilon \frac{n_S}{T_R^3}
\sim \epsilon \frac{T_R}{m} \, ,
\eeq
where $n_S$ is the number density of inflatons at the time of their decay.
This lepton (or $B-L$) asymmetry then generates a baryon
asymmetry \cite{fy,luty} through sphaleron interactions \cite{Manton:1983nd, Kuzmin:1985mm}.
The factor $\epsilon$ is a measure of the C and CP violation in the
decay, which is determined by loops in which one or both of the 
remaining singlet states is exchanged, and is given by \cite{Covi:1996wh}
\begin{equation}
 \epsilon \simeq -\frac{3}{8\pi}\frac{1}{\left(C_{SHL} C_{SHL}^\dagger \right)_{11}}
 \sum_{i=2,3} {\rm Im} \left[\left(C_{SHL}C_{SHL}^\dagger\right)^2_{1i}\right] 
\frac{M_1}{M_i} ~,
\end{equation} 
where $M_1 = m \ll M_2, M_3$ are the masses of the singlets,
where the lightest is assumed to be the inflaton~\footnote{See Ref.~\cite{ENO3} for a related discussion in the
context of flipped SU(5).}.
In order to obtain the correct baryon asymmetry, we should place additional constraints on the 
couplings and masses of the heavier singlets, which we do not discuss further here. 

\section{Summary}
\label{sec:summary}

It has been shown previously that no-scale supergravity with bilinear and trilinear
self-couplings of a singlet inflaton field provides an economical way to realize a model of inflation whose predictions
for the inflationary observables $(n_s, r)$ are similar to those of the Starobinsky model.
In this paper we have studied how this scenario may be embedded in a supersymmetric GUT
that is able to address other interesting phenomenological issues such as fermion (particularly
neutrino) masses, proton decay, leptogenesis, gravitino production and the nature of dark matter.

In this paper we have addressed these issues in a supersymmetric SO(10)
GUT model. In general, sneutrino inflation is an attractive scenario,
but this cannot be realized in an SO(10) GUT, because sneutrinos are
embedded in matter $\mathbf{16}$ representations of SO(10), but there
are no $\mathbf{16}^2$ or $\mathbf{16}^3$ couplings in SO(10). We
therefore consider an SO(10) GUT model with a singlet inflaton field, in
which there is an intermediate stage of symmetry breaking provided by a
Higgs $\mathbf{16}$ multiplet. This model has the K\"ahler potential
shown in (\ref{Kgen}) and the superpotential shown in (\ref{Wgen}).
As discussed in Section~\ref{sec:setup}, we consider various possible
patterns of symmetry breaking, paying careful attention to the vacuum
conditions in each case. 

We have shown that inflation can be realized in such a framework,
studying numerically the behaviours of the scalar fields during the
inflationary epoch. In particular, we tracked the evolution of the the
three Standard Model singlets in the {\bf 210} responsible for breaking
SO(10), the single in the Higgs {\bf 16} simultaneously with the
inflaton. One of the important phenomenological issues in constructing
such a GUT model is doublet-triplet mass splitting. As we have
discussed, the proton stability constraint requires either a very high
supersymmetry-braking scale and/or some additional mechanism to suppress
the color-triplet Higgs exchange contribution. These issues may be more
easily resolved in a flipped ${\rm SU}(5)\otimes {\rm U}(1)$ model
\cite{flipped2} where the Higgs structure is greatly simplified (only a
{\bf 10}, {$\mathbf{\overline{10}}$}, {\bf 5}, {$\mathbf{\bar{5}}$} of
Higgses are needed instead of the {\bf 210}, {\bf 16}, and
{$\mathbf{\overline{16}}$} considered here).

We have discussed the fermion masses in this model, point out that it predicts the (phenomenologically successful) 
unification of the $b$ and $\tau$ Yukawa couplings, and similar unification between the Yukawa couplings
in the up-type quark and neutrino sectors. The neutrino masses have a double-seesaw structure involving
the left- and right-handed neutrinos and the singlino partner of the inflaton field. We have explored the
constraints that neutrino masses impose on this structure, and shown that it can lead to successful
leptogenesis.

Two specifically supersymmetric issues are gravitino production during reheating at the end of inflation
and the nature of dark matter. Avoiding the overproduction of gravitinos imposes a reasonable constraint
on the inflaton Yukawa coupling, which should be at most comparable to that of the electron. In this model
$R$ parity is not conserved, so one might fear for the stability of supersymmetric dark matter. However,
the lifetime of the lightest supersymmetric particle is typically much longer than the age of the Universe,
so this is still a plausible candidate for dark matter.

The no-scale SO(10) GUT scenario for inflation described here has many attractive features, since it
combines Starobinsky-like predictions for the inflationary perturbations with many phenomenological
desiderata. We therefore consider it a significant step forward in inflationary model-building, while
admitting that it has some issues, notably proton stability. Thus there is still significant scope for
further improvement.

\section*{Acknowledgements}

The work of J.E. was supported in part by the UK STFC via the research grant ST/L000326/1.
The work of D.V.N. was supported in part by the DOE grant DE-FG02-13ER42020 and in part by the Alexander~S.~Onassis Public Benefit Foundation.
The work of M.A.G.G., N.N., and
K.A.O. was supported in part by DOE grant DE-SC0011842  at the University of
Minnesota.

\end{document}